\documentclass[aps,pra,twocolumn,floatfix]{revtex4}

\usepackage[dvips]{graphicx}
\usepackage{epsfig}
\usepackage{pst-plot}
\usepackage{bm}
\usepackage[latin1]{inputenc}
\usepackage{fontenc}
\usepackage{amsmath}
\usepackage{amssymb}
\usepackage{shadow}
\usepackage{pstricks}

\begin{document}
\title{Quantum computation algorithm for many-body studies}
\author{E.~Ovrum}
\affiliation{Department of Physics and 
Center of Mathematics for Applications, University of
  Oslo, 
N-0316 Oslo, Norway}
\author{M.~Hjorth-Jensen}
\affiliation{Department of Physics and Center of Mathematics for
  Applications, 
University of Oslo, N-0316 Oslo, Norway}
\date{\today}
\begin{abstract}
We show in detail how the Jordan-Wigner transformation can be  used to
simulate any fermionic many-body Hamiltonian on a quantum computer.
We develop an algorithm based on appropriate qubit gates that 
takes a general fermionic Hamiltonian, written as products of
a given number of creation and annihilation operators, as input.
To demonstrate the  applicability  of the algorithm, we calculate 
eigenvalues and eigenvectors of two model Hamiltonians, the
well-known
 Hubbard model and a generalized pairing Hamiltonian.  
Extensions to other systems are discussed.  
\end{abstract}

\maketitle

\section{Introduction}

A theoretical understanding of the behavior of many-body systems
is a  great challenge and provides fundamental insights into quantum 
mechanical studies, as well
as offering potential areas of applications.
However, apart from some few analytically solvable problems,
the typical absence of an exactly solvable contribution to the
many-particle
 Hamiltonian
means that we need reliable numerical many-body methods.
These methods should allow for controlled approximations
and provide a computational scheme which accounts for successive
many-body corrections in a systematic way.
Typical examples of
popular many-body methods are coupled-cluster methods
\cite{bartlett81,helgaker,Dean2004},
various types of
Monte Carlo methods \cite{Pudliner1997,kdl97,ceperley1995},
perturbative expansions \cite{lindgren,mhj95},
Green's function methods \cite{dickhoff,blaizot},
the density-matrix renormalization group \cite{white1992,schollwock2005},
ab initio density functional theory
\cite{bartlett2005,peirs2003,vanneck2006}
and large-scale diagonalization methods
\cite{Whitehead1977,caurier2005,navratil2004,horoi2006}. 

However, all these methods have to face in some form or the other the problem of 
an exponential growth in dimensionality. For a system of $P$ fermions
which 
can be placed 
into $N$ levels, the total number of basis states are given by
$\left(\begin{array}{c}N\\P\end{array}\right)$.
The dimensional curse means that most quantum
mechanical calculations on classical computers have exponential
complexity and therefore are very hard to solve for larger systems. On 
the other hand, a so-called 
quantum computer, a particularly dedicated computer,
can improve greatly on the size of systems that can be simulated, as
foreseen by Feynman \cite{feynman1982,feynman1986}. A quantum computer 
does not need
an exponential amount of memory to represent a quantum state.  
The basic unit of information for a  quantum computer 
is the so-called qubit or quantum bit. Any
suitable 
two-level quantum system can be a qubit, but the  
standard model of quantum computation is a model where two-level
quantum systems are located at different points in
space, and are manipulated by a small universal set of  operations.
These operations are called gates in the same fashion as operations on
bits in classical computers are called gates. 

For the example of $P$ spin $1/2$  particles, 
a classical computer needs $2^P$ bits to represent all possible states, while  
a quantum computer needs only $P$ qubits. The complexity in
number of qubits is thus linear.  Based on these ideas, several groups have proposed
various algorithms for simulating quantal many-body systems on quantum computers.
Abrams 
and  Lloyd, see for example Refs.~\cite{lloyd1997,lloyd1999a}, introduced a  
quantum algorithm that uses the quantum fast Fourier transform to find 
eigenvalues and eigenvectors of a given Hamiltonian,
illustrating how one could solve classically intractable problems with  
less than 100 qubits.
Achieving a polynomial complexity in the number of operations needed
to simulate a quantum system is not that straightforward however.
To get efficient simulations in time one
needs to transform the  many-body Hamiltonian into a sum of operations
on qubits, the building blocks of the quantum simulator and computer,
so that the time evolution operator can be implemented in polynomial
time. 
In Refs.~\cite{somma2002,somma2005,ortiz2002} it was shown how the
Jordan-Wigner 
transformation in principle
does this for all Hamiltonians acting on fermionic many-body states.
Based on this approach, recently two groups, see Refs.~\cite{krbrown2006,yang2006}, published 
results where 
they used Nuclear Magnetic Resonance (NMR) qubits to simulate the
pairing Hamiltonian.

The aim of this work is to develop an algorithm than allows one to perform   
a quantum computer simulation (or simply quantum simulation hereafter) of  any many-body
fermionic Hamiltonian. We show how to generate, via various Jordan-Wigner
transformations, all qubit operations
needed to simulate the time evolution operator of a given  Hamiltonian.
We also show that for a given term in an $m$-body fermionic
Hamiltonian, the number of operations needed to simulate it is linear
in the number of qubits or energy-levels of the system. The number of
terms in the Hamiltonian is of the order of $m^2$ for a
general $m$-body interaction, making the simulation increasingly
harder with higher order interactions. We specialize our examples to a two-body Hamiltonian,
since this is also the most general  type of Hamiltonian encountered in many-body physics.
Besides fields like nuclear physics, where three-body forces play a non-neglible role,
a two-body Hamiltonian captures most of the relevant physics. The various transformations are
detailed in the next section. In Sec.~\ref{sec:details} 
we show in detail how to simulate a quantum computer finding the
eigenvalues of any two-body Hamiltonian, with all available particle
numbers, 
using the simulated time
evolution operator. In that section we describe also the techniques which are necessary for the
extraction of information using a phase-estimation algorithm. 

To demonstrate the feasibility of our algorithm, 
we present in Sec.~\ref{sec:results} selected results from applications of 
our algorithm to two simple model-Hamiltonians, a pairing Hamiltonian and the Hubbard model.
We summarize our results and present future perspectives in  
Sec.~\ref{sec:conclusion}.

\section{Algorithm for quantum computations of fermionic systems}
\label{sec:algo}
\subsection{Hamiltonians}
A general two-body Hamiltonian for fermionic system 
can be written as 
\begin{equation}
\label{eq:twobodyH}
H = E_0 + \sum_{ij=1} E_{ij} a^\dag_i a_j
+\sum_{ijkl = 1} V_{ijkl} a^\dag_i a^\dag_j a_l a_k,
\end{equation} 
where $E_0$ is a constant energy term, $E_{ij}$ represent all the
one-particle terms, allowing for non-diagonal terms as well. The
one-body term can represent 
a chosen single-particle potential, the kinetic energy or other more
specialized terms such as  
those discussed in connection with the Hubbard model
\cite{hubbardmodel} or the pairing Hamiltonian  
discussed below.
The two-body interaction part is given by $V_{ijkl}$ and can be any
two-body interaction, from 
Coulomb interaction to the interaction between nucleons.  
The sums run over all possible single-particle levels $N$. 
Note that
this model includes particle numbers from zero to the number of
available quantum levels, $n$. To simulate states with fixed numbers
of fermions one would have to either rewrite the Hamiltonian or
generate specialized input states in the simulation.

The algorithm which we will develop in this section and in
Sec.~\ref{sec:details} can treat any 
two-body Hamiltonian. However, 
in our demonstrations of the quantum computing algorithm, we will limit ourselves to
two simple models, which however capture much of the important physics
in quantum mechanical 
many-body systems. We will also limit ourselves to spin $j=1/2$
systems, although our algorithm 
can also simulate  higher $j$-values, such as those which occur in nuclear, atomic and
molecular physics, it simply uses one qubit for every available
quantum state.   
These simple models are the Hubbard model and a pairing
Hamiltonian.
We start with the spin $1/2$ Hubbard model, described by the following Hamiltonian
\begin{eqnarray}
H_H &&= \epsilon \sum_{i, \sigma} a_{i\sigma}^\dag a_{i\sigma} 
-t \sum_{i, \sigma} \left(a^\dag_{i+1, \sigma}a_{i, \sigma}
+a^\dag_{i, \sigma}a_{i+1, \sigma} \right) \notag \\
&& + U \sum_{i=1} a_{i+}^\dag a_{i-}^\dag a_{i-}a_{i+},
\label{eq:hubbard}
\end{eqnarray}
where $a^{\dagger}$ and $a$ are fermion creation and annihilation operators, respectively.
This is a chain of sites where each site has room for one spin up
fermion and one spin down fermion. 
The number of sites is $N$, and the sums over $\sigma$ are sums over
spin up and down only.
Each site has a single-particle
energy $\epsilon$. There is a repulsive term $U$ if there is a pair
of particles at the same site. It is energetically favourable to tunnel to
neighbouring sites, described by 
the hopping terms with coupling constant $-t$.

The second model-Hamiltonian is the simple pairing Hamiltonian 
\begin{equation}
   H_P=\sum_i \varepsilon_i a^{\dagger}_i a_i -\frac{1}{2} g\sum_{ij>0}
           a^{\dagger}_{i}
     a^{\dagger}_{\bar{\imath}}a_{\bar{\jmath}}a_{j},
     \label{eq:pairing}
\end{equation}
 The indices $i$ and $j$ run over the number of levels $N$, and the label $\bar{\imath}$ 
stands for a time-reversed state. The parameter $g$ is the strength of the pairing force 
while $\varepsilon_i$ is the single-particle energy of level $i$. 
In our case
we assume that the single-particle levels are equidistant (or
degenerate) with a fixed spacing $d$. 
Moreover, in our simple model, the degeneracy of the single-particle
levels is set to $2j+1=2$, with $j=1/2$  
being the spin of the particle. This gives a set of single-particle
states with the same spin projections as 
for the Hubbard model.  Whereas in the Hubbard model we operate with
different sites with  
spin up or spin down particles, our pairing models deals thus with
levels  with double degeneracy. 
Introducing the pair-creation operator 
$S^+_i=a^{\dagger}_{im}a^{\dagger}_{i-m}$,
one can rewrite the Hamiltonian in 
Eq.\ (\ref{eq:pairing}) as
\[
   H_P=d\sum_iiN_i+
     \frac{1}{2} G\sum_{ij>0}S^+_iS^-_j,
\]
where  $N_i=a^{\dagger}_i a_i$
is the number operator, and 
$\varepsilon_i = id$ so that the single-particle orbitals 
are equally spaced at intervals $d$. The latter commutes with the 
Hamiltonian $H$. In this model, quantum numbers like seniority 
$\cal{S}$ are good quantum numbers, and the eigenvalue problem 
can be rewritten in terms of blocks with good seniority. 
Loosely 
speaking, the seniority quantum number $\cal{S}$ is equal to 
the number of unpaired particles; see  \cite{Talmi1993} for 
further details. 
Furthermore, in  a series of papers, Richardson, see for example 
Refs.~\cite{richardson1,richardson2,richardson3},  
obtained the exact solution of the pairing Hamiltonian, with 
semi-analytic (since there is still the need for a numerical solution) 
expressions for the eigenvalues and eigenvectors. The exact solutions
have had important consequences for several fields, from Bose condensates to
nuclear superconductivity and is currently a very active field of studies, see for example
Refs.~\cite{dukelsky2004,rmp75mhj}.
Finally, for particle numbers up to $P \sim 20$, the above model can be 
solved exactly through numerical diagonalization and one can obtain all eigenvalues.
It serves therefore also as an excellent ground for comparison with our algorithm based
on models from quantum computing.

\subsection{Basic quantum  gates}

Benioff showed that one could make a quantum mechanical Turing machine
by using various  unitary operations on a quantum system, see Ref.~\cite{benioff}.
Benioff  demonstrated 
that a quantum computer can calculate anything a
classical computer can. To do this one needs a quantum system and
basic operations that can approximate all unitary operations
on the chosen many-body system. We describe in this subsection the basic ingredients entering 
our algorithms.

\subsubsection{Qubits, gates and circuits}
\label{sec:gates}
In this article we will use the standard model of quantum information,
where
the basic unit of information is the qubit, the quantum bit. 
As mentioned in the introduction, any
suitable 
two-level quantum system can be a qubit, 
it is the smallest system there is with the
least complex dynamics.
Qubits are both abstract measures of information and physical objects.
Actual physical qubits can be ions trapped in magnetic fields where
lasers can access only two energy levels or  the nuclear spins of some of
the atoms in molecules accessed and manipulated by an NMR machine.
Several other ideas have been proposed and some tested, see
\cite{nielsen2000}. 

The computational basis for one qubit is ${\ensuremath{|0\rangle}}$ (representing for example bit $0$) 
for the first state
and ${\ensuremath{|1\rangle}}$ (representing bit $1$) for the second, and for a set of qubits  
the tensor products of
these basis states for each qubit form a product basis. Below we write out the different
basis states for a system of $n$ qubits.
\begin{eqnarray}
\label{eq:compBasis}
&{\ensuremath{|0\rangle}} \equiv {\ensuremath{|00\cdots 0\rangle}} =
      {\ensuremath{|0\rangle}} \otimes {\ensuremath{| 0\rangle}} \otimes
          \cdots 
\otimes {\ensuremath{|0\rangle}} 
\notag \\
&{\ensuremath{|1\rangle}} \equiv {\ensuremath{|00\cdots 1\rangle}} =
    {\ensuremath{|0\rangle}} \otimes {\ensuremath{| 0\rangle}} \otimes
        \cdots 
\otimes {\ensuremath{|1\rangle}} 
\notag \\
&\vdots \notag \\
&{\ensuremath{|2^n-1\rangle}} \equiv {\ensuremath{|11\cdots 1\rangle}} =
    {\ensuremath{|1\rangle}} \otimes {\ensuremath{| 1\rangle}} \otimes 
\cdots \otimes {\ensuremath{|1 \rangle}}.
\notag \\
\end{eqnarray}
This is a $2^n$-dimensional system and we number the different basis
states using binary numbers corresponding to the order in which they appear in the
tensor product.

Quantum computing means to  manipulate and measure  qubits in such a
way that the results from a measurement yield the solutions to  a given problem. 
The quantum operations we need to be able to perform our simulations are 
a small set of elementary single-qubit
operations, or single-qubit gates, and one universal two-qubit gate,
in our case the so-called CNOT gate defined below.

\begin{figure}[h]
\begin{picture}(0,65)(160,55)
\put(60,87){\makebox(0,0){${\ensuremath{| a\rangle}}$}}
\put(60,63){\makebox(0,0){${\ensuremath{| b\rangle}}$}}
\put(90,50){\input{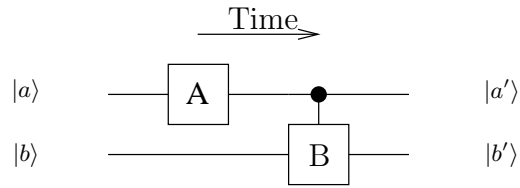}}
\put(240, 87){\makebox(0,0){${\ensuremath{|{a^\prime\rangle}}}$}}
\put(240, 63){\makebox(0,0){${\ensuremath{| {b^\prime\rangle}}}$}}
\end{picture}
\caption{A quantum circuit showing a single-qubit gate $A$ and a 
two-qubit gate acting on a pair of qubits,  represented by the horizontal lines.}
\label{fig:circ}
\end{figure}

\begin{figure}[h!]
\begin{picture}(200,30)
\scalebox{1.1}{\put(10,10){\makebox(0,0){phase}}}
\put(50,-2){\input{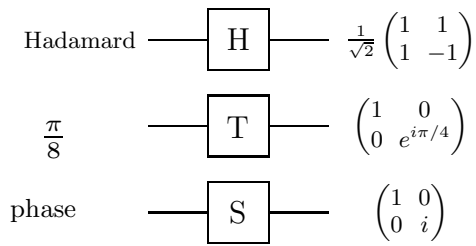}}
\put(150, 10){\makebox(0,0){\ensuremath{
	\begin{pmatrix}
	1 & 0 \\
	0 & i \end{pmatrix}
    }}}
\end{picture}
\caption{Set of three elementary single-qubit gates and their matrix representations. 
The products of these three
operations on one qubit can approximate to an arbitrary precision any
unitary operation on that qubit.}
\label{fig:elementarySingleQubitGates}
\end{figure}

To represent quantum computer algorithms graphically we use circuit
diagrams. In a circuit diagram each qubit is represented by a line,
and operations on the different qubits are represented by boxes.
In fig.~\ref{fig:circ} we show an example of a quantum circuit, with the arrow 
indicating the time evolution,
The states ${\ensuremath{| a\rangle}}$ and ${\ensuremath{| b\rangle}}$ in the figure represent qubit 
states. In general, the total state will be a 
superposition of different qubit states.
A single-qubit gate is an operation that only affects one physical
qubit, for example one ion or one nuclear spin in a molecule. It is represented by
a box on the line corresponding to the qubit in question. A
single-qubit 
gate operates on one qubit and is therefore represented
mathematically by a $2\times2$ matrix while a two-qubit gate is
represented by a $4\times4$ matrix. As an example we can portray the so-called CNOT
gate as matrix,
\begin{equation}
	\begin{pmatrix}
	1 & 0 & 0 & 0\\
	0&1&0&0\\
	0&0&0&1\\
	0&0&1&0
	\end{pmatrix}.
\label{eq:CNOT}
\end{equation}
This is a very important gate, since one can show that it behaves as a universal two-qubit
gate, and that we only need this two-qubit gate and a small set of single-qubit
gates to be able to approximate any multi-qubit operation. One example of a
universal set of single-qubit gates is given in
Fig.~\ref{fig:elementarySingleQubitGates}. The products of these three
operations on one qubit can approximate to an arbitrary precision any
unitary operation on that qubit.
\begin{figure}[h]
\begin{picture}(200,20)
\put(25,-10){\makebox(0,0){Hadamard}}
\put(50,-22){\input{fig2.latex}}
\put(150, -10){\makebox(0,0){\ensuremath{
	\frac{1}{\sqrt2}
	\begin{pmatrix}
	1 & 1 \\
	1 &-1 \end{pmatrix}
    }}}
\end{picture}
\end{figure}

\begin{figure}[h]
\begin{picture}(200,10)
\scalebox{1.5}{\put(10,-10){\makebox(0,0){$\frac\pi8$}}}
\put(50,-22){\input{fig3.latex}}
\put(150, -10){\makebox(0,0){\ensuremath{
	\begin{pmatrix}
	1 & 0 \\
	0 &e^{i\pi/4} \end{pmatrix}
    }}}
\end{picture}
\end{figure}

\subsubsection{Decomposing unitary operations into gates}
The next step is to find elementary operations
on a set of qubits that can be put together in order to approximate any unitary
operation on the qubits. In this way we can perform computations on
a quantum computer by performing many of these elementary operations
in the correct order.

There are three steps in finding the elementary operations needed to
simulate any unitary operation.
First,  any
$d\times d$ unitary matrix can be factorized into a product of at most
$d(d-1)/2$ two-level unitary matrices, see for example Ref.~\cite{nielsen2000} for details. 
A two-level unitary matrix is a
matrix that only acts non-trivially on two vector components when
multiplied with a vector. For all other vector components it acts as
the identity operation. 

The next step is to prove that any two-level unitary matrix can be
implemented by one kind of two-qubit gate, for example the CNOT gate 
in Eq.~(\ref{eq:CNOT}), 
and single-qubit gates only. 
This simplifies the making of actual quantum
computers as we only need one type of interaction between pairs of
qubits. All other operations are operations on one qubit at the time. 

Finally,  these single-qubit operations can be approximated to an
arbitrary precision by a finite set of single-qubit gates. Such a set
is called a universal set and one example is the phase gate, the
so-called Hadamard gate and
the $\pi/8$ gate. Fig.~\ref{fig:elementarySingleQubitGates} shows these gates. 
By combining these properly 
with the CNOT gate one can approximate any unitary operation on a set
of qubits. 

\subsubsection{Quantum calculations}
The aspect of quantum computers we are focusing on in this article is
their use in computing properties of quantum systems. 
When we want to use a quantum computer to 
find the energy levels of a quantum system
or simulate it's dynamics, we need to simulate
the time evolution operator of the Hamiltonian, $U=\exp(-iH\Delta t)$.
To do that on a quantum computer we must find a set of single- and 
two-qubit gates that would implement the time evolution on a set of
qubits. For example, if we have one qubit in the state ${\ensuremath{|
    a\rangle}}$, we must find the single-qubit gates that when applied
results in the qubit being in the state
$\exp(-iH\Delta t)|a\rangle$. 

From what we have written so far
the naive way of simulating $U$ would be to calculate it directly as
a matrix in an appropriate basis, factorize it into two-level unitary
matrices and then implement these by a set of universal gates. 
In a many-body fermion system for example, one could use the Fock basis to
calculate $U$ as a matrix.  A fermion system with $n$ different
quantum levels can have from zero to $n$ particles in each Fock basis
state. A 
two-level system has four different basis states, $|00\rangle$,
$|01\rangle$, $|10\rangle$ and $|11\rangle$, where $|0\rangle$ corresponds to an
occupied quantum level.  The time evolution matrix is then a $2^n\times
2^n$ matrix.  This matrix is then factorized into at most
$2^n(2^n-1)/2$ two-level unitary matrices. An exponential
amount of operations, in terms of the number of quantum levels, is
needed to simulate $U$; by definition not an effective simulation.

This shows that quantum computers performing quantum
simulations not necessarily fulfill their promise. For each physical
system to be simulated one has to find  representations of the
Hamiltonian that leads to polynomial complexity in the number of
operations. After one has found a proper representation of the
Hamiltonian, the time evolution operator $\exp(-iH\Delta t)$ is found by using a 
Trotter approximation, for example 
\begin{equation}
\label{eq:Trotter1}
U=e^{-iH\Delta t}=e^{-i(\sum_k H_k) \Delta t} = \prod_k e^{-iH_k\Delta
  t} + {\cal O}(\Delta t^2).
\end{equation}
There are different ways to approximate $U$ by products of 
exponentials of the different terms of the Hamiltonian, see
Ref.~\cite{nielsen2000} and Eq.~(\ref{eq:Trotter2}). The essential 
idea is to find a form of the Hamiltonian where these factors in the
approximated time evolution operator can be further
factorized into single- and two-qubit operations effectively. 
In
Refs.~\cite{ortiz2001,ortiz2002}
it was shown how to do this in principle for
any many-body fermion system using the Jordan-Wigner transformation.
In this article we 
show how to create a quantum  compiler that takes
any many-body fermion Hamiltonian and outputs the quantum gates needed
to simulate the time evolution operator. We implement it for the case
of two-body fermion Hamiltonians and show results from numerical
calculations finding the energylevels of the well known pairing and
Hubbard models.

\subsection{The Jordan-Wigner transformation}

For a spin-$1/2$ one-dimensional quantum spin-chain a fermionization
procedure exists 
which allows the mapping between spin operators and fermionic
creation-annihilation operators.
The algebra governing the spin chain is the $SU(2)$ algebra, represented by
the $\sigma$-matrices. The Jordan-Wigner transformation is a
transformation from fermionic annihilation and creation operators to
the $\sigma$-matrices of a spin-$1/2$ chain, see
for example Ref.~\cite{dargis1998} for more details on the Jordan-Wigner
transformation. 

There is an isomorphism 
between the two systems, meaning that any $a$ or $a^\dag$ operator can be
transformed into a tensor product of $\sigma$-matrices operating on a
set of qubits. This was explored by Somma {\em et al.} in Refs.~\cite{somma2002,ortiz2002}.  
The authors demonstrated, with an emphasis on single-particle fermionic operators,  
that the Jordan-Wigner transformation
ensures efficient, i.e., not exponential complexity, simulations of a
fermionic system on a quantum computer. 
Similar transformations must be found for other systems, in order to
efficiently simulate many-body systems. This was the main
point in Ref.~\cite{somma2002}. 

We present here the various ingredients needed in order to transform a given 
Hamiltonian into a practical form suitable  for quantum mechanical simulations. 

We begin with the  fermionic creation and annihilation operators, which satisfy the following
anticommutation relations
\begin{equation}
\label{eq:anticommutationrelations}
\{a_k, a_l\}=\{a_k^\dag, a_l^\dag\}= 0, \quad 
\{a_k^\dag, a_l\} = \delta_{kl}.
\end{equation}
Thereafter we define the three traceless and Hermitian generators of the $SU(2)$ group, the
$\sigma$-matrices $\sigma_x$, $\sigma_y$ and $\sigma_z$.  Together 
with the
identity matrix ${\bf 1}$ they form a complete basis for all Hermitian $2\times2$
matrices. They can be used to write all Hamiltonians on a spin $1/2$
chain when taking sums of tensor products of these, in other words  
they form a product basis for the operators on the qubits.
The three $\sigma$-matrices are
\begin{equation}
\sigma_x = \begin{pmatrix}
  0 & 1\\
  1 & 0 
\end{pmatrix},
\sigma_y = \begin{pmatrix}
  0 & -i\\
  i & 0 
\end{pmatrix}, \quad
\sigma_z = \begin{pmatrix}
  1 & 0\\
  0 & -1 
\end{pmatrix}.
\end{equation}
We define the raising and lowering matrices as
\[
\sigma_+ = \frac{1}{2}(\sigma_x + i\sigma_y)=
\begin{pmatrix}
  0 & 1\\
  0 & 0 
\end{pmatrix},
\]
\begin{equation}
\label{eq:raisingAndLowerin}
\sigma_- = \frac{1}{2}(\sigma_x - i\sigma_y)
=\begin{pmatrix}
  0 & 0\\
  1 & 0 
\end{pmatrix}.
\end{equation}
The transformation is based on the fact that for each possible
quantum state of the fermion system, there can be either one or zero
fermions. Therefore we need $n$ qubits for a system with $n$ possible
fermion states. A qubit in state $ |0\rangle ^i=a^\dag_i|vacuum\rangle$ 
represents a state with a
fermion, while $ |1\rangle ^i=|vacuum\rangle$ represents no fermions. Then the raising
operator $\sigma_+$ changes $ |1\rangle $ into $ |0\rangle $ when
\begin{equation}
 |0\rangle  \equiv \begin{pmatrix} 1 \\ 0 \end{pmatrix}, \quad
 |1\rangle  \equiv \begin{pmatrix} 0 \\ 1 \end{pmatrix}.
\end{equation}  
This means that $\sigma_+$ acts as a creation operator, and $\sigma_-$
acts as an annihilation operator. In addition, because of the
anticommutation of creation(annihilation)  operators for different states we have 
$a_1^\dag a_2^\dag
|vacuum\rangle = - a_2^\dag a_1^\dag |vacuum\rangle$, meaning that for creation and
annihilation operators for states higher than the state corresponding
to the first qubit, we need to multiply with a $\sigma_z$-matrix on
all the qubits leading up to the one in question, in order  to get the correct
sign in the final operation. This leads us to the Jordan-Wigner
transformation \cite{somma2002,ortiz2002}
\begin{equation}
\label{eq:JWtransformation}
a^\dag_n = \left(\prod_{k=1}^{n-1} \sigma_z^k\right) \sigma_+^n, \quad 
a_n = \left(\prod_{k=1}^{n-1} \sigma_z^k\right) \sigma_-^n.
\end{equation}
The notation $\sigma_z^i\sigma_+^j$ means a tensor product of the
identity matrix on all qubits other than $i$ and $j$, ${\bf 1}\otimes
\sigma_z \otimes {\bf 1} \otimes \sigma_+\otimes{\bf 1}$, if $i<j$, with ${\bf 1}$
being the  identity matrices of appropriate dimension.

\subsection{Single-particle Hamiltonian}
\label{sec:1partH}
What we must do now is to apply the Jordan-Wigner transformation to a
general fermionic Hamiltonian composed of creation and annihilation
operators, so we can write it as a sum of products of $\sigma$
matrices. The matrix $\sigma^k$ is then an operation on the $k^{th}$
qubit representing the $k^{th}$ quantum level of the fermion system.
When we have expressed the Hamiltonian as a sum of products of
operations on the qubits representing the system, we must find a
representation of the time evolution operator as products of two-qubit
operations. These operations can be further decomposed into elementary
operations, see subsection \ref{sec:gates} for further discussion.

\subsubsection{Jordan-Wigner transformation of the one-body part}
We first describe the procedure for the 
simplest case of a general single-particle Hamiltonian,
\begin{equation}
H_1=\sum_{i} E_{ii} a^\dag_i a_i +
\sum_{i<j} E_{ij} (a^\dag_i a_j + a^\dag_j a_i).
\end{equation} 
We now  use the transformation  of  
Eq.~(\ref{eq:JWtransformation}) on the terms $a^\dag_i a_j$. 

The diagonal terms of the one-particle Hamiltonian, 
that is the case where $i=j$, can be rewritten as
\[
a^\dag_i a_i = \left(\prod_{k=1}^{i-1} \sigma_z^k\right) \sigma_+^i
\left(\prod_{k=1}^{i-1} \sigma_z^k \right)\sigma_-^i
\]
\begin{equation}
\label{eq:H_1Sigmas}
= \sigma_+^i \sigma_-^i = \frac{1}{2}\left({\bf 1}^i + \sigma_z^i\right),
\end{equation}
since $(\sigma_z)^2= {\bf 1}$ which is the number operator. It counts whether or not a fermion is in
state $i$. In the case of qubits counting whether or not the qubit
is in state $ |0\rangle $, we have eigenvalue one for $ |0\rangle $ and eigenvalue
zero for $ |1\rangle $. The action of this Hamiltonian on qubit $i$ can be simulated using
the single-qubit operation 
\begin{equation}
U=e^{-i({\bf 1} + \sigma_z) E_{ii}\Delta t} = \begin{pmatrix}
  e^{-iE_{ii}\Delta t} & 0 \\
  0 & 1 
\end{pmatrix},
\end{equation}
see subsection \ref{sec:gates} for other examples of single-qubit gates.

For the non-diagonal elements, $i<j$,  not all of the $\sigma_z$ matrices multiply with each
other and end up in the identity operation. 
As an example we will consider a five level system, $n=5$, and look
at the transformation of the term $a^\dag_ia_j$ whith $i=2$ and $j=4$,
\begin{eqnarray}
\label{eq:fiveQubitSystem}
a_2^\dag& = &\sigma_z\otimes\sigma_+\otimes {\bf 1} \otimes {\bf 1}\otimes {\bf 1}, \notag\\
a_4 & = & \sigma_z\otimes\sigma_z\otimes\sigma_z\otimes\sigma_-\otimes {\bf 1}, \notag\\ 
 & \Downarrow & \notag\\
a_2^\dag a_4 & = & {\bf 1}\otimes (\sigma_+\sigma_z)\otimes \sigma_z \otimes \sigma_-\otimes {\bf 1}.
\end{eqnarray}
The operation on all qubits
before $i$ and after $j$ is identity, on qubits $i+1$ through
$j-1$ it is  $\sigma_z$.  
We can then write the non-diagonal one-body operators as
\begin{widetext}
\begin{eqnarray}
\label{eq:singleParticle}
a^\dag_i a_j + a^\dag_j a_i &&= (\sigma_+^i\sigma_z^i)
\left(\prod_{k=i+1}^{j-1} 
\sigma_z^k \right)\sigma_-^j
+
(\sigma_z^i\sigma_-^i) \left(\prod_{k=i+1}^{j-1}
\sigma_z^k \right)\sigma_+^j \notag\\
&&= - \sigma_+^i \left(\prod_{k=i+1}^{j-1}\sigma_z^k \right)\sigma_-^j
- \sigma_-^i \left(\prod_{k=i+1}^{j-1}
\sigma_z^k \right)\sigma_+^j \notag\\
&&= -\frac{1}{2} \left\{ 
\sigma_x^i \left(\prod_{k=i+1}^{j-1}\sigma_z^k \right)\sigma_x^j
+ \sigma_y^i \left(\prod_{k=i+1}^{j-1}\sigma_z^k \right)\sigma_y^j
\right\}.
\end{eqnarray}
\end{widetext}

Using Eqs.~(\ref{eq:H_1Sigmas}) and (\ref{eq:singleParticle})
the total single-particle fermionic Hamiltonian of $n$ quantum levels,
transformed using the Jordan-Wigner transformation of
Eq.~(\ref{eq:JWtransformation}), is written as
\begin{eqnarray}
\label{eq:singleParticleTotal}
H_1 &=& \sum_{i} E_{ii} a^\dag_i a_i +
\sum_{i<j} E_{ij} (a^\dag_i a_j + a^\dag_j a_i) \notag\\
&=&
 \frac{1}{2}\sum_{i} E_{ii}\left({\bf 1}^i + \sigma_z^i\right)
\notag\\
&-& \frac{1}{2}\sum_{i<j} E_{ij} 
  \left\{ 
\sigma_x^i \left(\prod_{k=i+1}^{j-1}\sigma_z^k \right)\sigma_x^j \right.\notag\\
&+& \left.\sigma_y^i \left(\prod_{k=i+1}^{j-1}\sigma_z^k \right)\sigma_y^j
\right\}.
\end{eqnarray}

\subsubsection{Transformation into two-qubit operations}
The Hamiltonian is now transformed into a sum of many-qubit
operations, $H=\sum_l H_l$. The $a_2^\dag a_4$ term in Eq.~(\ref{eq:fiveQubitSystem})
for example, is transformed into a three-qubit operation. The next
step is to factorize these many-qubit operations $H_l$ into products of
two-qubit operations, so that we in the end can get a product of
two-qubit operations $U_{kl}$, that when performed in order give us the time
evolution operator corresponding to each term in the Hamiltonian, 
$\exp(-iH_l\Delta t) = \prod_k U_{kl}$.

The first thing we do is to find a set of two-qubit operations that
together give us the Hamiltonian, and later  we will see that to
find the time evolution from there is straightforward. 
The many-qubit terms in Eq.~(\ref{eq:singleParticleTotal}) are 
products of the type $\sigma_x \sigma_z \cdots \sigma_z \sigma_x$
with $\sigma_x$ or $\sigma_y$ at either end. These products have to be
factorized into a series of two-qubit operations.
What we wish to do is successively build up the operator using
different unitary transformations. 
This can be achieved with successive operations with the $\sigma$-matrices, starting with for example
$\sigma_z^i$, which can be transformed 
into $\sigma_x^i$, then transformed into $\sigma_y^i\sigma_z^{i+1}$
and so forth.
Our goal now is to express each term in the Hamiltonian
Eq.~(\ref{eq:singleParticleTotal}) as a product of the type
$\sigma_x^i \sigma_z \cdots \sigma_z 
\sigma_x^j=
(\prod_k U_k^\dag ) \sigma_z^i 
(\prod_{k^\prime} U_{k^\prime})$, with a different form in the case
where the Hamiltonian term starts and ends with a $\sigma_y$ matrix.
To achieve this we need the transformations in
Eqs.~(\ref{eq:rotations1})-(\ref{eq:rotations4}). 
We will use this to find the time-evolution operator for each Hamiltonian, see
Eq.~(\ref{eq:Us}) below. 

To understand how we factorize the Hamiltonian terms into single- and
two-qubit operations we follow a bottom up procedure.
First, 
if we have a two qubit system, with the operator $\sigma_z \otimes
{\bf 1}$, we see that the unitary operation $\exp(i\pi/4 \sigma_z\otimes
\sigma_z)$  transforms it into
\begin{equation}
e^{-i\pi/4 \sigma_z\otimes\sigma_z}\left( \sigma_z\otimes{\bf 1} \right)
e^{i\pi/4 \sigma_z\otimes\sigma_z} = \sigma_z\otimes\sigma_z.
\end{equation} 
In addition, if we start out with the operator $\sigma_z^i$ we can transform it
into $\sigma_x^i$ or $\sigma_y^i$ using the operators
$\exp(i\pi/4\sigma_y)$ or $\exp(-i\pi/4\sigma_x)$ accordingly. 

We can then generate the $\prod_k \sigma_z^k$ part of the terms in
Eq.~(\ref{eq:singleParticleTotal}) by succesively
applying the operator $\exp(i\pi/4\sigma_z^i \sigma_z^l)$ for $l=2$
through $l=j$. Yielding the operator $\sigma_a^i \prod_{k=i+1}^{j}
\sigma_z^k$ with a phase of $\pm1$, because of the sign change in
Eqs.~(\ref{eq:rotations3}) and (\ref{eq:rotations4}). We write
$\sigma_a$ to show that we can start with both a $\sigma_x$ and
a $\sigma_y$ matrix. To avoid the sign
change we can simply use the operator $\exp(-i\pi/4\sigma_z^i
\sigma_z^l)$ instead for those cases where  we have $\sigma_y^i$ on site
$i$ 
instead of $\sigma_x^i$. This way we always keep the same phase.

Finally, we use the operator $\exp(i\pi/4\sigma_y)$ if we want the
string of operators to end with $\sigma_x$, or  $\exp(-i\pi/4\sigma_x)$
if we want it to end with $\sigma_y$.
The string of operators starts with either $\sigma_x$ or $\sigma_y$. For an odd
number of $\exp(\pm i\pi/4\sigma_z^i \sigma_z^l)$ operations, the
operations that add a $\sigma_z$ to the string, the first
operator has changed from what we started with. In other words we have
$\sigma_x$ instead of $\sigma_y$ at the start of the string or vice
versa, see Eqs.~(\ref{eq:rotations3}) and (\ref{eq:rotations4}).
By counting, we see that we do 
$j-i$ of the  $\exp(\pm i\pi/4\sigma_z^i \sigma_z^l)$
operations to get the string
$\sigma_a^i\sigma_z^{i+1}\cdots\sigma_z^j$. 
and therefore if $j-i$ is odd, the first matrix is the opposite of
what we want in the final string.
The following simple example can serve to clarify.
We want the Hamiltonian
$\sigma_x^1\sigma_z^2\sigma_x^3= \sigma_x\otimes \sigma_z \otimes
\sigma_x$, and by using the transformations in
Eqs.~(\ref{eq:rotations1})-(\ref{eq:rotations4}) we can construct it
as a product of single- and two-qubit operations in the following way,
\begin{eqnarray}
(e^{-\pi/4\sigma_y^1}) \sigma_z^1 (e^{\pi/4\sigma_y^1})&=&\sigma_x^1
\notag\\
 (e^{-i\pi/4\sigma_z^1\sigma_z^2}) \sigma_x^1
 (e^{i\pi/4\sigma_z^1\sigma_z^2}) & = & \sigma_y^1\sigma_z^2 \notag \\
(e^{i\pi/4\sigma_z^1\sigma_z^3}) \sigma_y^1\sigma_z^2
(e^{-i\pi/4\sigma_z^1\sigma_z^3}) & = & \sigma_x^1\sigma_z^2 \sigma_z^3
\notag\\
(e^{-i\pi/4\sigma_y^3}) \sigma_x^1\sigma_z^2 \sigma_z^3
(e^{i\pi/4\sigma_y^3}) & =& \sigma_x^1\sigma_z^2\sigma_x^3.
\end{eqnarray}
We see that we have factorized $\sigma_x^1\sigma_z^2\sigma_x^3$ into
$U_4^\dag U_3^\dag U_2^\dag U_1^\dag \sigma_z^1 U_1 U_2 U_3U_4$.

Now we can find  
the time-evolution operator $\exp(-iH\Delta t)$ corresponding to each
term of the Hamiltonian, which is the quantity of interest. 
Instead of starting
with the operator $\sigma_z^i$ we start with the corresponding
evolution operator and observe that
\begin{eqnarray}
U^\dag e^{-i\sigma_z a} U &&= U^\dag \left( \cos(a){\bf 1}
-i\sin(a) \sigma_z \right) U\notag \\
&&=\cos(a){\bf 1} -i \sin(a) U^\dag \sigma_zU \notag \\
&&=e^{-i U^\dag\sigma_zU a},
\end{eqnarray}
where $a$ is  a scalar.
This means that we have a series of unitary transformations on this
operator yielding the final evolution,  namely
\begin{equation}
\label{eq:Us}
e^{-i \sigma_x^i \sigma_z \cdots \sigma_z \sigma_x^j a} 
= \left(\prod_k U_k^\dag \right) e^{-i\sigma_z^ia} 
\left(\prod_{k^\prime}
U_{k^\prime}\right),
\end{equation}
with the exact same unitary operations $U_k$ as we find when we
factorize the Hamiltonian. These are now the single- and two-qubit operations we
were looking for, first we apply the operations $U_k$ to the
appropriate qubits, then $\exp(-i\sigma_z^ia)$ to qubit $i$, and then
the $U_k^\dag$ operations, all in usual matrix multiplication order.

\subsection{Two-body Hamiltonian}
\label{sec:2bH}
In this section we will do the same for the general two-body fermionic
Hamiltonian.
The two-body part of the Hamiltonian can be classified into
diagonal elements and non-diagonal elements. Because of the Pauli
principle and the anti-commutation relations for the creation and
annihilation operators, some combinations of indices are not allowed.
The two-body part of our Hamiltonian is 
\begin{equation}
H_2 = \sum_{ijkl} V_{ijkl} a_i^\dag a_j^\dag a_l a_k,
\end{equation} 
where the indices run over all possible states and $n$  is the total number
of available quantum states. 
The single-particle labels $ijkl$ refer to their 
corresponding sets of quantum numbers, such as 
projection of total spin, number of nodes in the single-particle wave function etc. Since every state
$ijkl$ is uniquely defined,
we cannot have two equal
creation or annihilation operators and therefore $i\neq j$ and $k\neq
l$.

When $i=l$ and $j=k$, or $i=k$ and $j=l$, we have a
diagonal element in the Hamiltonian matrix, and the output state
is the same as the input state. 
The operator term corresponding to $V_{ijji}$ has these equalities due
to the anti-commutation relations
\begin{eqnarray}
a^\dag_i a^\dag_j a_i a_j &&=  a^\dag_j a^\dag_i a_j a_i \notag\\
&&= -a^\dag_i a^\dag_j a_j a_i \notag\\
&&= -a^\dag_j a^\dag_i a_i a_j,
\end{eqnarray}
which means that
\begin{equation}
\label{eq:Vdiags}
V_{ijji} = V_{jiij} = - V_{ijij} = - V_{jiji}.
\end{equation}
The term $a^\dag_i a^\dag_j a_i a_j$
with $i<j$ is described using the Pauli matrices
\begin{eqnarray}
&&a^\dag_i a^\dag_j a_i a_j \\ &&= 
\left(\prod_{s=1}^{i-1} \sigma_z \right) \sigma_+^i 
\left(\prod_{t=1}^{j-i} \sigma_z \right)
\sigma_+^j \notag \\
&&\times\left(\prod_{t=1}^{j-i} \sigma_z \right) \sigma_-^j
\left(\prod_{s=1}^{i-1} \sigma_z \right) \sigma_-^i \notag \\
&&=\left(\prod_{s=1}^{i-1} (\sigma_z)^4 \right) \left( \sigma_+^i \sigma_z^i
\sigma_z^i \sigma_-^i\right) \left(\prod_{t=i+1}^{j-1} (\sigma_z)^2 \right) \left(
\sigma_+^j  
\sigma_-^j\right) \notag \\
&&=\sigma_+^i\sigma_-^i\sigma_+^j\sigma_-^j \notag\\
&&=\frac{1}{16} \left( {\bf 1} + \sigma_z^i\right) \left( {\bf 1} + \sigma_z^j
\right). 
\end{eqnarray}
When we add all four different permutations of
$i$ and $j$ this is the number operator on qubit $i$ multiplied with
the number operator on qubit $j$.
The eigenvalue is one if both qubits are in
the state $ |0\rangle $, that is the corresponding quantum states are both
populated, and zero otherwise.
We can in turn rewrite the sets of creation and annihilations in terms of the 
$\sigma$-matrices as
\begin{eqnarray}
a^\dag_i a^\dag_j a_i a_j + a^\dag_j a^\dag_i a_j a_i
-a^\dag_i a^\dag_j a_j a_i-a^\dag_j a^\dag_i a_i a_j \notag \\
= \frac{1}{4} \left( {\bf 1} 
+\sigma_z^i + \sigma_z^j +\sigma_z^i\sigma_z^j \right).
\end{eqnarray}

In the general case we can have three different sets of non-equal
indices. Firstly, we see that $a^\dag_i a^\dag_j a_l a_k = a^\dag_k
a^\dag_l a_j a_i$, meaning that the  exchange of $i$ with $k$ and $j$ with $l$
gives the same operator $\rightarrow V_{ijkl}=V_{klij}$. This results in 
a two-body Hamiltonian with no equal indices
\begin{equation}
\label{eq:H2nonequalSymmetric}
H_{ijkl} = 
\sum_{i < k}\sum_{ j < l} V_{ijkl} (a_i^\dag a_j^\dag a_l a_k + a_k^\dag
a_l^\dag a_j a_i).
\end{equation} 
Choosing to order the 
indices from lowest to highest gives us  the position 
where there will be $\sigma_z$-matrices to multiply with the
different raising and lowering operators, when we perform the
Jordan-Wigner transformation Eq.~(\ref{eq:JWtransformation}). The
order of matrix 
multiplications is fixed once and for all, resulting in  three
different groups into which these terms fall, namely
\begin{equation}
\label{eq:threeGroups}
\begin{array}{ccccc}
I & i<j<l<k, & i \leftrightarrow j, &  k\leftrightarrow l, \\
II & i<l<j<k, \quad& i \leftrightarrow j, &  k\leftrightarrow l, \\
III & i<l<k<j, \quad& i \leftrightarrow j, &  k\leftrightarrow l. \\
\end{array} 
\end{equation}
These $12$ possibilities for $a^\dag_i a^\dag_j a_l a_k$ 
are mirrored in the symmetric term
in Eq.~(\ref{eq:H2nonequalSymmetric}) giving us the $24$ different
possibilities when permuting four indices.

The $ijkl$ term of Eq.~(\ref{eq:H2nonequalSymmetric}) is 
\begin{eqnarray}
a^\dag_i a^\dag_j a_l a_k + a_k^\dag
a_l^\dag a_j a_i= 
\notag\\
\left(\prod \sigma_z \right) \sigma_+^i \left(\prod \sigma_z \right)
\sigma_+^j \notag \\
\times\left(\prod \sigma_z \right) \sigma_-^l
\left(\prod \sigma_z \right) \sigma_-^k \notag \\
+\left(\prod \sigma_z \right) \sigma_+^k \left(\prod \sigma_z \right)
\sigma_+^l \notag \\
\times
\left(\prod \sigma_z \right) \sigma_-^j
\left(\prod \sigma_z \right) \sigma_-^i .
\end{eqnarray}
In the case of $i<j<l<k$ we have
\begin{eqnarray}
\label{eq:VijklPhi}
a^\dag_i a^\dag_j a_l a_k+ a_k^\dag a_l^\dag a_j a_i=\notag\\
\left(\prod (\sigma_z)^4\right) \left(\sigma_+^i \sigma_z^i\right)
\left(\prod (\sigma_z)^3\right) \sigma_+^j \notag \\
\times
\left(\prod (\sigma_z)^2\right) \left(\sigma_-^l \sigma_z^l \right)
\left(\prod \sigma_z\right) \sigma_-^k \notag \\
+\left(\prod (\sigma_z)^4\right) \left(\sigma_z^i\sigma_-^i\right)
\left(\prod (\sigma_z)^3\right) \sigma_-^j \notag \\
\times
\left(\prod (\sigma_z)^2\right) \left(\sigma_z^l\sigma_+^l \right)
\left(\prod \sigma_z\right) \sigma_+^k .
\end{eqnarray}
Using Eq.~(\ref{eq:pmzs}), where  we have the rules for sign changes when
multiplying the raising and lowering operators with the $\sigma_z$
matrices, gives us
\begin{eqnarray}
-\left(\sigma_+^i \sigma_z^{i+1}\cdots \sigma_z^{j-1} \sigma_+^j
\sigma_-^l \sigma_z^{l+1}\cdots \sigma_z^{k-1}\sigma_-^k\right. 
\notag\\
+ \left.\sigma_-^i \sigma_z^{i+1}\cdots \sigma_z^{j-1} \sigma_-^j
\sigma_+^l \sigma_z^{l+1}\cdots \sigma_z^{k-1}\sigma_+^k\right).
\end{eqnarray}

If we switch the order of $i$ and $j$ so that $j<i<l<k$, we change the
order in which the $\sigma_z$-matrix is multiplied with the first
raising and lowering matrices, resulting in a sign change.
\begin{eqnarray}
a^\dag_i a^\dag_j a_l a_k  + a_k^\dag a_l^\dag a_j a_i
=\notag\\
\left(\prod (\sigma_z)^4\right) \left(\sigma_z^j \sigma_+^j\right)
\left(\prod (\sigma_z)^3\right) \sigma_+^i \notag \\
\times 
\left(\prod (\sigma_z)^2\right) \left(\sigma_-^l \sigma_z^l \right)
\left(\prod \sigma_z\right) \sigma_-^k \notag \\
 +\left(\prod (\sigma_z)^4\right) \left(\sigma_-^j\sigma_z^j\right)
\left(\prod (\sigma_z)^3\right) \sigma_-^i \notag \\
\times 
\left(\prod (\sigma_z)^2\right) \left(\sigma_z^l\sigma_+^l \right)
\left(\prod \sigma_z\right) \sigma_+^k \notag \\
=+\left(\sigma_+^j \sigma_z^{j+1}\cdots \sigma_z^{i-1} \sigma_+^i
\sigma_-^l \sigma_z^{l+1}\cdots \sigma_z^{k-1}\sigma_-^k\right.
\notag\\
 \left.+ \sigma_-^j \sigma_z^{j+1}\cdots \sigma_z^{i-1} \sigma_-^i
\sigma_+^l \sigma_z^{l+1}\cdots \sigma_z^{k-1}\sigma_+^k\right).
\end{eqnarray}

We get a change in sign for every permutation of the ordering
of the indices from lowest to highest because of the matrix
multiplication ordering. The ordering is described by another set of
indices \newline $\{s_\alpha, s_\beta, s_\gamma, s_\delta\} \in \{i, j, k, l\}$
where  
$s_\alpha< s_\beta< s_\gamma< s_\delta$. We assign a number to each of
the four indices, $i\leftrightarrow 1$, $j\leftrightarrow 2$,
$l\leftrightarrow 3$ and $k\leftrightarrow 4$. If $i<j<l<k$ we say the
ordering is $\alpha =1$, $\beta =2$, $\gamma=3$ and $\delta=4$, where
$\alpha$ is a number from one to four indicating which of the indices $i$, $j$,
$l$ and $k$ is the smallest. If $i$ is the smallest, $\alpha=1$ and
$s_\alpha= i$. This allows us to give the sign of a
given $(a^\dag_i a^\dag_j a_l a_k + a_k^\dag a_l^\dag a_j a_i)$ 
term using the totally anti-symmetric tensor with
four indices, which is $+1$ for even permutations, and $-1$ for odd
permutations. For each of the three groups in Eq.~(\ref{eq:threeGroups})
we get a different set of raising and lowering operators on the
lowest, next lowest and so on, indices, while the sign for the whole
set is given by $-\varepsilon^{\alpha\beta\gamma\delta}$.

We are in the position where we can use the relation in 
Eq.~(\ref{eq:raisingAndLowerin}) to express
the Hamiltonian in terms of the $\sigma$-matrices.  
We get 16 terms with products of four
$\sigma_x$ and or $\sigma_y$ matrices in the first part of
Eq.~(\ref{eq:VijklPhi}),
then when we add the Hermitian conjugate we get another 16 terms.
The terms with an odd number of $\sigma_y$ matrices have an imaginary
phase and are therefore cancelled out
when adding the conjugates in Eq.~(\ref{eq:H2nonequalSymmetric}).
This leaves us with just the terms with four $\sigma_x$
matrices, four $\sigma_y$ matrices and two of each in different
orderings. 
The final result is given as an array with a global sign and factor
given by the permutation of the ordering, and eight terms with
different signs depending on which of the three groups,
Eq.~(\ref{eq:threeGroups}), the set of indices belong to.
These differing rules are due to the rules for $\sigma_z$
multiplication with the raising and lowering operators,  resulting  in 
\begin{eqnarray}
\label{eq:Phiijkl}
&&a^\dag_i a^\dag_j a_l a_k + a_k^\dag a_l^\dag a_j a_i= \notag \\
&&-\frac{\varepsilon^{\alpha\beta\gamma\delta}}{8} \left\{
\begin{array}{cccc}
I & II & III &  \\
+& + & + & \sigma_x^{s_\alpha} \sigma_z \cdots \sigma_z 
\sigma_x^{s_\beta} \sigma_x^{s_\gamma}\sigma_z\cdots \sigma_z
\sigma_x^{s_\delta}\\
-& +& +& \sigma_x \cdots \sigma_x \quad \sigma_y \cdots\sigma_y \\
+& -& +& \sigma_x \cdots \sigma_y \quad \sigma_x \cdots\sigma_y \\
+& +& -& \sigma_x \cdots \sigma_y \quad \sigma_y \cdots\sigma_x \\
+& +& -& \sigma_y \cdots \sigma_x \quad \sigma_x \cdots\sigma_y \\
+& -& +& \sigma_y \cdots \sigma_x \quad \sigma_y \cdots\sigma_x \\
-& +& +& \sigma_y \cdots \sigma_y \quad \sigma_x \cdots\sigma_x \\
+& +& +& \sigma_y \cdots \sigma_y \quad \sigma_y \cdots\sigma_y \\
\end{array} 
\right.
\end{eqnarray}
where the letters $I$, $II$ and $III$ refer to the subgroups defined
in Eq.~(\ref{eq:threeGroups}). 

As for the single-particle operators in subsection \ref{sec:1partH}
we now need to factorize these multi-qubit 
terms in the Hamiltonian to
products of two-qubit and single-qubit operators.
Instead of transforming a product of the form $a z\cdots z b$, we now
need to transform a product of the form $a z\cdots zb cz\cdots zd$,
where $a$, $b$, $c$ and $d$ are short for either $\sigma_x$ or
$\sigma_y$ while $z$ is short for $\sigma_z$.  
The generalization is quite straightforward, as we see that if the
initial operator is $\sigma_z^{s_\alpha} \sigma_z^{s_\gamma}$ instead
of just $\sigma_z^{s_\alpha}$, we can use the same set of
transformations as for the single-particle case,
\begin{eqnarray}
&&U_k^\dag \cdots U_1^\dag \sigma_z^{s_\alpha} U_1 \cdots U_k 
\notag  \\ &&=
\sigma_a^{s_\alpha} 
\sigma_z \cdots \sigma_z \sigma_b^{\beta}
\notag \\
\Rightarrow&&  
U_k^\dag \cdots U_1^\dag \sigma_z^{s_\alpha} \sigma_z^{s_\gamma} 
U_1 \cdots U_k \notag \\
&&= \sigma_a^{s_\alpha}
\sigma_z \cdots \sigma_z \sigma_b^{s_\beta}  \sigma_z^{s_\gamma}.
\end{eqnarray}
Using the same unitary two-qubit transformations, which we now call 
$V$, 
that take $\sigma_z^{s_\gamma}$ to $\sigma_c^{s_\gamma} \sigma_z
\cdots \sigma_z \sigma_d^{s_\delta}$,
we find
\begin{eqnarray}
&&V_s^\dag\cdots V_1^\dag
U_k^\dag \cdots U_1^\dag \sigma_z^{s_\alpha}\sigma_z^{s_\gamma}
 U_1 \cdots U_k 
V_1\cdots V_s \notag \\
&&=
\sigma_a^{s_\alpha} 
\sigma_z \cdots \sigma_z \sigma_b^{\beta} 
\sigma_c^{s_\gamma} 
\sigma_z \cdots \sigma_z \sigma_d^{\delta}.
\end{eqnarray}
This straightforward generalization of the procedure from the single-particle
Hamiltonian case is possible because the operations commute when performed
on different qubits.

With the above expressions, we can start with
the unitary operator $\exp(-ia \sigma_z^{s_\alpha}\sigma_z^{s_\gamma})$
and have two different series of unitary operators that give us the
evolution operator of the desired Hamiltonian. 
The $U$ operators are defined as in 
Eq.~(\ref{eq:Us}),
\begin{equation}
e^{-i\sigma^{s_\alpha} \sigma_z \cdots \sigma_z
\sigma^{s_\beta}a}  
= \left(\prod_k U_k^\dag \right) e^{-i\sigma_z^{s_\alpha}a} 
\left(\prod_{k^\prime}
U_{k^\prime}\right),
\end{equation}
while the 
$V$ operators are defined in a similar way
\begin{equation}
e^{-i\sigma^{s_\gamma} \sigma_z \cdots \sigma_z
\sigma^{s_\delta}a}  
= \left(\prod_s V_s^\dag \right) e^{-i\sigma_z^{s_\gamma}a} 
\left(\prod_{s^\prime}
V_{s^\prime}\right),
\end{equation}
where the $\sigma$-matrices without subscripts represent that we can
have $\sigma_x$ or $\sigma_y$ in each position.

This gives us the total evolution operator for each term in
Eq.~(\ref{eq:Phiijkl})  
\begin{eqnarray}
&&e^{-i\sigma^{s_\alpha} \sigma_z \cdots \sigma_z
\sigma^{s_\beta} \sigma^{s_\gamma} \sigma_z \cdots \sigma_z
\sigma^{s_\delta} a} \notag \\
&&=\left(\prod_s V_s^\dag \right)\left(\prod_k U_k^\dag \right)
e^{-i \sigma_z^{s_\alpha}\sigma_z^{s_\gamma}a} \notag\\
&&\times\left(\prod_{k^\prime}
U_{k^\prime}\right)
\left(\prod_{s^\prime}
V_{s^\prime}\right).
\end{eqnarray}
Here we have all the single- and two-qubit operations we need to
perform on our set of qubits, that were initially in the state
$|\psi\rangle$, to simulate the time evolution $\exp(-iH_k\Delta
t)|\psi\rangle$ of the Hamiltonian term $H_k= \sigma^{s_\alpha} \sigma_z \cdots \sigma_z
\sigma^{s_\beta} \sigma^{s_\gamma} \sigma_z \cdots \sigma_z
\sigma^{s_\delta}$. Every factor in the above equation is a single- or
two-qubit operation that must be performed on the qubits in proper
matrix multiplication order.

When using the Jordan-Wigner transformation of
Eq.~(\ref{eq:JWtransformation}) applied to our two model Hamiltonians of
Eqs.~(\ref{eq:hubbard}) and (\ref{eq:pairing}), we choose a representation with two qubits at each site. These
correspond to fermions with spin up and down, respectively.
The number of qubits, $n$, is always the total number of available
quantum states and therefore
it is straightforward to use this model on systems with higher degeneracy, such
as those encountered in quantum chemistry \cite{helgaker} or nuclear physics \cite{caurier2005}. 
Site one spin up is qubit one, site one spin
down is qubit two and site two spin up is qubit three and so on.
To get all the quantum gates one needs to simulate a given Hamiltonian
one needs to input the correct $E_{ij}$ and $V_{ijkl}$ values. 

\subsection{Complexity of the quantum computing algorithm}
\label{sec:complexityOfFermionicSimulator}

In order to test the efficiency of a quantum algorithm, one needs to know how many
qubits, and how many operations on these, are  needed to implement the
algorithm.  Usually this is a 
function of the dimension of the Hilbert space on which the
Hamiltonian acts. The natural input scale 
in the fermionic simulator is the number of quantum states, $n$, that are
available to the fermions.
In our simulations  of the Hubbard and the pairing
models of Eqs.~(\ref{eq:hubbard}) and (\ref{eq:pairing}), respectively, 
the number of qubits is $n=2N$ since we have chosen systems with 
double-degeneracy for every single-particle state, where $N$ is the
number of energy-levels in the model.
We use one qubit to represent each possible fermion state,
on a real quantum computer, however, one should implement some
error-correction procedure using several qubits for each state, see
Ref.~\cite{nielsen2000}. 
The complexity in number of qubits remains linear, however, since ${\cal O} (n)$
qubits are needed for error correction.

The single-particle Hamiltonian has  potentially ${\cal O} (n^2)$
different $E_{ij}$ terms. The two-particle Hamiltonian has up to
${\cal O} (n^4)$ $V_{ijkl}$ terms. A general $m$-body interaction has
in the worst case  ${\cal O} (n^{2m})$ terms. It is straightforward to
convince oneself that the
pairing model has  ${\cal O} (n^2)$ terms  while
in the Hubbard model we end up with   ${\cal O}
(n)$ terms. Not all models have maximum complexity in the different
$m$-body interactions.

How many two-qubit operations do each of these terms need to be
simulated? First of all a two-qubit operation will in general have to
be decomposed into a series of universal single- and two-qubit
operations, depending entirely on the given quantum simulator. A particular
physical realization might have a natural implementation of the
$\sigma_z^i\otimes \sigma_z^j$ gate and save a lot of intermediary
operations.
Others will have to use a fixed number of operations in order to apply  the
operation on any two qubits. A system with only nearest neighbor
interactions would have to use ${\cal O}(n)$ operations for each
$\sigma_z^i\otimes \sigma_z^j$ gate, and thereby increase the
polynomial complexity by one degree. 

In our discussion on the one-body part of the Hamiltonian, 
we saw that for each $E_{ij}$ we obtained the
$a^\dag_ia_j + a^\dag_j a_i$ operator which is transformed into the
two terms in Eq.~(\ref{eq:singleParticle}), 
$\sigma_x\sigma_z\cdots\sigma_z \sigma_x$ and 
$\sigma_y\sigma_z\cdots\sigma_z \sigma_y$. 
We  showed how these terms are
decomposed into $j-i+2$ operations, leading to twice as many 
unitary transformations on an operator, $VAV^\dag$ for the time evolution.
The average of $j-i$ is $n/2$ in this case and in total we need to
perform $2\times 2\times n/2 = 2n$ two-qubit operations per
single-particle term in the Hamiltonian, a linear complexity.

In the two-particle case each term $V_{ijkl}(a^\dag_i a^\dag_j a_l a_k
+ a^\dag_k a^\dag_l a_j a_i)$ is transformed into a sum of eight
operators of the form $\sigma^{s_\alpha} \sigma_z \cdots \sigma_z
\sigma^{s_\beta} \sigma^{s_\gamma} \sigma_z \cdots \sigma_z
\sigma^{s_\delta}$, Eq.~(\ref{eq:Phiijkl}). The two parts of these
operators are implemented in the same way as the $\sigma^i
\sigma_z\cdots \sigma_z \sigma^j$ term of the single-particle
Hamiltonian, which means they require $s_\beta - s_\alpha$  and
$s_\delta -s_\gamma$ operations, since
$s_\alpha<s_\beta<s_\gamma<s_\delta$  the average is $n/4$. 
For both of these parts we need to perform both the unitary operation
$V$ and it's
Hermitian conjugate $V^\dag$. In the end we need $2\times 2
\times 8 \times n/4=8n$ two-qubit operations per two-particle term in
the Hamiltonian, the complexity is linear.

A term of an $m$-body Hamiltonian will be transformed into $2^{2m}$
operators since each annihilation and creation operator is transformed
into a sum of $\sigma_x$ and $\sigma_y$ matrices. 
All the imaginary terms cancel out and we are left with $2^{2m-1}$
terms. 
Each of these terms
will include $2m$ $\sigma$ matrices, in products of the form
$\prod_{k=1}^m \sigma^i \sigma_z \cdots \sigma_z \sigma^j$,
and we use the same procedure as discussed above 
to decompose these $m$  factors into
unitary 
transformations. In this case each factor will require an average of
$n/2m$ operations for the same reasons as in the two-body case. 
All in all, each $m$-body term in the Hamiltonian requires
$2^{2m-1}\times 2\times m \times n/2m = 2^{2m-1}n$ operations.

Thus, the complexity for simulating one $m$-body term of a fermionic
many-body  Hamiltonian is linear in the number of two-qubit
operations, but the number of terms is  not. For a
full-fledged simulation of general three-body forces, in common use in
nuclear physics \cite{Pieper2001,navratil2002,ccsdt03}, 
the total complexity of the simulation is ${\cal O}
(n^7)$.  A complete two-particle Hamiltonian would be ${\cal O}
(n^5)$.The bottleneck in these simulations is the number of terms in the
Hamiltonian, and for systems with less than the full number of terms
the simulation will be faster.
This is much better than the exponential complexity of
most simulations on classical computers.

\section{Algorithmic details}
\label{sec:details}
Having detailed how a general Hamiltonian, of two-body nature in our case,
can be decomposed in terms of various quantum gates,we present here details of the 
implementation of our algorithm for finding eigenvalues and eigenvectors of
a many-fermion system.
For our tests of the fermionic simulation algorithm we have
implemented the phase-estimation algorithm from \cite{nielsen2000}
which finds the eigenvalues of an Hamiltonian operating on a set of
simulation qubits.  There are also other quantum computer algorithms for
finding expectation values and correlation functions, as discussed by Somma {\em et al.} in 
Refs.~\cite{somma2002,somma2005}. 
In the following we first describe the phase-estimation algorithm, and
then describe its implementation  and methods we have developed
in using this algorithm. A much more thorough description of quantum
computers and the phase-estimation algorithm can be found in
\cite{ovrum2003}.

\subsection{Phase-estimation algorithm}

To find the eigenvalues of the Hamiltonian we use
the unitary time evolution operator we get from the Hamiltonian. 
We have a set of simulation qubits representing the system governed by
the Hamiltonian, and a set of auxiliary qubits, called work qubits \cite{lloyd1997,lloyd1999a}, in
which we will store the eigenvalues of the time evolution operator.
The procedure is to perform several controlled time evolutions with
work 
qubits as control qubits and the simulation qubits as targets,  
see for example Ref.~\cite{nielsen2000} for information on controlled qubit
operations. 
For
each work qubit we perform the controlled operation on the simulation
qubits with a different time parameter, giving all the work qubits
different phases. The information stored in their phases is extracted
using first an inverse Fourier transform on the work qubits alone, and
then performing a measurement on them. The values of the
measurements give us directly the eigenvalues of the Hamiltonian after
the algorithm has been performed a number of times.

The input state of the simulation qubits is a random state in our
implementation, which is
also a random superposition of the eigenvectors of the Hamiltonian
$|\psi\rangle=\sum_k c_k | k\rangle$. It does not have to be a random state,
and in \cite{lawu2002} the authors describe a quasi-adiabatic
approach, where the initial state is created by starting in the ground 
state for the non-interacting Hamiltonian, 
a qubit basis state, e.g. $|0101\cdots101\rangle$, 
and
then slowly the interacting part of the Hamiltonian is turned on.
This gives us an initial state mostly comprising the true ground state,
but it can also have parts of the lower excited states if the
interacting Hamiltonian is turned on a bit faster.
In for example nuclear physics it is common to use a starting state for large-scale
diagonalizations that reflects some of the features of the states one wishes to study.
A typical example is to include pairing correlations in the trial wave function, see
for example Refs.~\cite{caurier2005,rmp75mhj}.
Iterative methods such as the Lanczo's diagonalization technique  \cite{Whitehead1977,golub1996}
converge much faster if such starting vectors are used. However, although more iterations are needed,
even a random starting vector converges to the wanted states.
 
The final state of all the qubits
after an inverse Fourier transform on the work qubits is 
\begin{equation}
\label{eq:FinalState}
	\sum_{k}c_k |\phi^{[k]} 2^t\rangle \otimes |k\rangle.
\end{equation}
If the
algorithm works perfectly, $|k\rangle$  should be an exact eigenstate of
$U$, with an exact eigenvalue $\phi^{[k]}$.
When we have the 
eigenvalues of the time evolution operator we easily find 
the eigenvalues of the
Hamiltonian. 
We can summarize schematically the phase-estimation algorithm as follows:
\begin{enumerate}
\item Intialize each of the work qubits to $1/\sqrt2 ( |0\rangle  + |1\rangle )$
  by initializing to $ |0\rangle $ and applying the Hadamard gate, H,
  see Fig.~\ref{fig:elementarySingleQubitGates}.
\item Initialize the simulation qubits to a random or  specified
  state, depending on the whether one wants the whole eigenvalue
  spectrum. 
\item  Perform  conditional time evolutions on the simulation qubits, with
  different timesteps $\Delta t$ and different work qubits as the
  control qubits.
\item Perform an inverse Fourier transform on the work qubits.
\item Measure the work qubits to extract the phase.
\item Repeat steps 1-6 until the probability distribution gathered
  from the measurement results is good enough to read out the wanted
  eigenvalues. 
\end{enumerate} 
\begin{widetext}
\onecolumngrid 
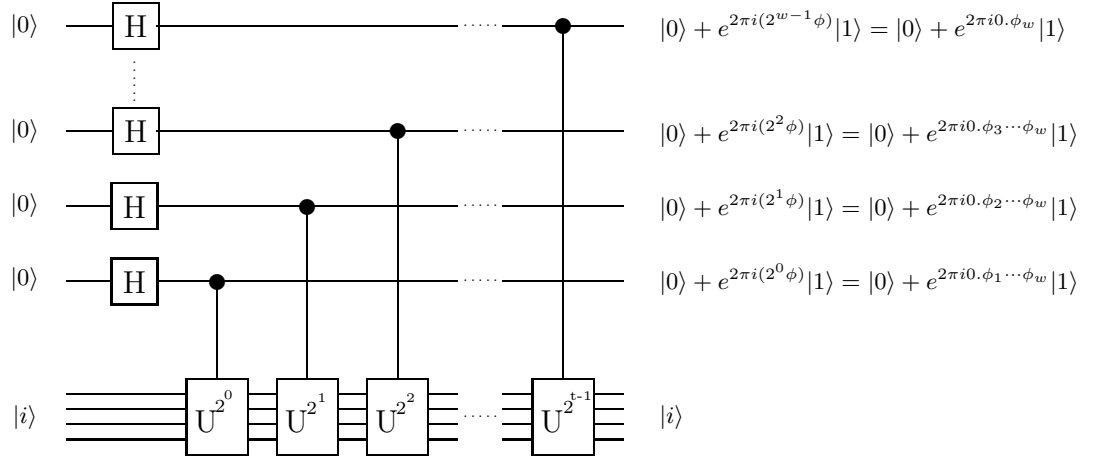
\begin{figure}
\begin{picture}(250,200)
\put(15,0){\setlength{\unitlength}{4144sp}%
\begingroup\makeatletter\ifx\SetFigFont\undefined%
\gdef\SetFigFont#1#2#3#4#5{%
  \reset@font\fontsize{#1}{#2pt}%
  \fontfamily{#3}\fontseries{#4}\fontshape{#5}%
  \selectfont}%
\fi\endgroup%
\begin{picture}(3354,2724)(529,-2413)
\thinlines
\put(819, 29){\framebox(270,270){}}
\put(886, 97){\makebox(0,0)[lb]{\smash{\SetFigFont{12}{14.4}{\rmdefault}{\mddefault}{\updefault}H}}}
\put(819,-601){\framebox(270,270){}}
\put(886,-533){\makebox(0,0)[lb]{\smash{\SetFigFont{12}{14.4}{\rmdefault}{\mddefault}{\updefault}H}}}
\put(811,-1051){\framebox(270,270){}}
\put(878,-983){\makebox(0,0)[lb]{\smash{\SetFigFont{12}{14.4}{\rmdefault}{\mddefault}{\updefault}H}}}
\put(811,-1501){\framebox(270,270){}}
\put(878,-1433){\makebox(0,0)[lb]{\smash{\SetFigFont{12}{14.4}{\rmdefault}{\mddefault}{\updefault}H}}}
\put(1441,-1366){\circle*{90}}
\put(1981,-916){\circle*{90}}
\put(2521,-466){\circle*{90}}
\put(3511,164){\circle*{90}}
\multiput(946,-61)(0.00000,-45.00000){7}{\makebox(1.5875,11.1125){\SetFigFont{5}{6}{\rmdefault}{\mddefault}{\updefault}.}}
\put(541,164){\line( 1, 0){270}}
\put(541,-466){\line( 1, 0){270}}
\put(541,-916){\line( 1, 0){270}}
\put(541,-1366){\line( 1, 0){270}}
\put(541,-2041){\line( 1, 0){720}}
\put(541,-2131){\line( 1, 0){720}}
\put(541,-2221){\line( 1, 0){720}}
\put(541,-2311){\line( 1, 0){720}}
\put(1261,-2401){\framebox(360,450){}}
\put(1621,-2041){\line( 1, 0){180}}
\put(1621,-2131){\line( 1, 0){180}}
\put(1621,-2221){\line( 1, 0){180}}
\put(1621,-2311){\line( 1, 0){180}}
\put(1801,-2401){\framebox(360,450){}}
\put(2161,-2041){\line( 1, 0){180}}
\put(2161,-2131){\line( 1, 0){180}}
\put(2161,-2221){\line( 1, 0){180}}
\put(2161,-2311){\line( 1, 0){180}}
\put(2341,-2401){\framebox(360,450){}}
\put(2701,-2041){\line( 1, 0){180}}
\put(2701,-2131){\line( 1, 0){180}}
\put(2701,-2311){\line( 1, 0){180}}
\put(2701,-2221){\line( 1, 0){180}}
\multiput(2926,-2176)(45.00000,0.00000){5}{\makebox(1.5875,11.1125){\SetFigFont{5}{6}{\rmdefault}{\mddefault}{\updefault}.}}
\put(3151,-2041){\line( 1, 0){180}}
\put(3151,-2131){\line( 1, 0){180}}
\put(3151,-2221){\line( 1, 0){180}}
\put(3151,-2311){\line( 1, 0){180}}
\put(3331,-2401){\framebox(360,450){}}
\put(3691,-2041){\line( 1, 0){180}}
\put(3691,-2131){\line( 1, 0){180}}
\put(3691,-2221){\line( 1, 0){180}}
\put(3691,-2311){\line( 1, 0){180}}
\put(1081,-1366){\line( 1, 0){1800}}
\put(1081,-916){\line( 1, 0){1800}}
\put(1081,-466){\line( 1, 0){1800}}
\put(1081,164){\line( 1, 0){1800}}
\multiput(2926,-1366)(45.00000,0.00000){5}{\makebox(1.5875,11.1125){\SetFigFont{5}{6}{\rmdefault}{\mddefault}{\updefault}.}}
\multiput(2926,-916)(45.00000,0.00000){5}{\makebox(1.5875,11.1125){\SetFigFont{5}{6}{\rmdefault}{\mddefault}{\updefault}.}}
\multiput(2926,-466)(45.00000,0.00000){5}{\makebox(1.5875,11.1125){\SetFigFont{5}{6}{\rmdefault}{\mddefault}{\updefault}.}}
\multiput(2926,164)(45.00000,0.00000){5}{\makebox(1.5875,11.1125){\SetFigFont{5}{6}{\rmdefault}{\mddefault}{\updefault}.}}
\put(3151,-1366){\line( 1, 0){720}}
\put(3151,-916){\line( 1, 0){720}}
\put(3151,-466){\line( 1, 0){720}}
\put(3151,164){\line( 1, 0){720}}
\put(1441,-1366){\line( 0,-1){585}}
\put(1981,-916){\line( 0,-1){1035}}
\put(2521,-466){\line( 0,-1){1485}}
\put(3511,164){\line( 0,-1){2115}}
\put(1306,-2266){\makebox(0,0)[lb]{\smash{\SetFigFont{12}{14.4}{\rmdefault}{\mddefault}{\updefault}U}}}
\put(1846,-2266){\makebox(0,0)[lb]{\smash{\SetFigFont{12}{14.4}{\rmdefault}{\mddefault}{\updefault}U}}}
\put(2386,-2266){\makebox(0,0)[lb]{\smash{\SetFigFont{12}{14.4}{\rmdefault}{\mddefault}{\updefault}U}}}
\put(3376,-2266){\makebox(0,0)[lb]{\smash{\SetFigFont{12}{14.4}{\rmdefault}{\mddefault}{\updefault}U}}}
\put(1426,-2146){\makebox(0,0)[lb]{\smash{\SetFigFont{7}{8.4}{\rmdefault}{\mddefault}{\updefault}2}}}
\put(1976,-2161){\makebox(0,0)[lb]{\smash{\SetFigFont{7}{8.4}{\rmdefault}{\mddefault}{\updefault}2}}}
\put(2511,-2161){\makebox(0,0)[lb]{\smash{\SetFigFont{7}{8.4}{\rmdefault}{\mddefault}{\updefault}2}}}
\put(3511,-2151){\makebox(0,0)[lb]{\smash{\SetFigFont{7}{8.4}{\rmdefault}{\mddefault}{\updefault}2}}}
\put(1481,-2071){\makebox(0,0)[lb]{\smash{\SetFigFont{5}{6.0}{\rmdefault}{\mddefault}{\updefault}0}}}
\put(2041,-2086){\makebox(0,0)[lb]{\smash{\SetFigFont{5}{6.0}{\rmdefault}{\mddefault}{\updefault}1}}}
\put(2571,-2086){\makebox(0,0)[lb]{\smash{\SetFigFont{5}{6.0}{\rmdefault}{\mddefault}{\updefault}2}}}
\put(3551,-2076){\makebox(0,0)[lb]{\smash{\SetFigFont{5}{6.0}{\rmdefault}{\mddefault}{\updefault}t-1}}}
\end{picture}}
\put(0,15){\makebox(0,0){\ensuremath{|i\rangle}}}
\put(0,67){\makebox(0,0){\ensuremath{|0\rangle}}}
\put(0,95){\makebox(0,0){\ensuremath{|0\rangle}}}
\put(0,123){\makebox(0,0){\ensuremath{|0\rangle}}}
\put(0,163){\makebox(0,0){\ensuremath{| 0\rangle}}}
\put(240,15){\makebox(0,0)[l]{\ensuremath{|i\rangle}}}
\put(240,67){\makebox(0,0)[l]{\ensuremath{ |0\rangle  + e^{2\pi
i(2^0\phi)} |1\rangle  =  |0\rangle  + e^{2\pi i0.\phi_1\cdots\phi_w} |1\rangle }}}
\put(240,95){\makebox(0,0)[l]{\ensuremath{ |0\rangle  + e^{2\pi i(2^1\phi)} |1\rangle 
 	= |0\rangle  + e^{2\pi i0.\phi_2\cdots\phi_w} |1\rangle }}}
\put(240,123){\makebox(0,0)[l]{\ensuremath{ |0\rangle  + e^{2\pi i(2^2\phi)} |1\rangle =
	 |0\rangle  + e^{2\pi i 0.\phi_3\cdots\phi_w} |1\rangle }}}
\put(240,163){\makebox(0,0)[l]{\ensuremath{ |0\rangle  + e^{2\pi
i(2^{w-1}\phi)} |1\rangle 
	= |0\rangle  + e^{2\pi i0.\phi_w} |1\rangle }}}
\end{picture}
\caption{Phase estimation circuit showing all the different qubit
  lines schematically with operations represented by boxes. The boxes
  connected by vertical lines to other qubit lines are controlled
  operations, with the qubit with the black dot as the control qubit.}
\label{fig:phaseEstCircuit}
\end{figure}
\end{widetext}
\twocolumngrid

As discussed above a set of two-qubit
operations can be simulated by the CNOT two-qubit operation and a
universal set of single-qubit operations. 
We will not use or discuss any such
implementation in this article, as one will have to use a different
set for each physical realization of a quantum computer. When
simulating a fermion system with a given
quantum computer, our algorithm  will first take the
fermionic many-body evolution operator to a series of two-qubit and
single-qubit operations, and then one will have to have a system
dependent setup that takes these operations to the basic building
blocks that form the appropriate universal set.

In subsection \ref{sec:2bH} we showed how to take any two-particle
fermionic Hamiltonian to a set of two-qubit operations that
approximate 
the evolution operator.
In addition we must use one of the Trotter approximations  \cite{trotter1959,suzukitrotter,suzuki1985}
Eqs.~(\ref{eq:Trotter1}) and  (\ref{eq:Trotter2})  that take
the evolution operator of a sum of terms to the product of the
evolution operator of the individual terms,
see for example Ref.~\cite{nielsen2000} for details. To order ${\cal O}(\Delta
t^2)$ in the error  we use Eq.~(\ref{eq:Trotter1})
while to order ${\cal O} (\Delta t^3)$ we have
\begin{equation}
	\label{eq:Trotter2} 
	e^{-i(A+B)\Delta t}=e^{-iA\Delta
	t/2}e^{-iB\Delta t}e^{-iA\Delta t/2} + {\cal O} (\Delta t^3).
\end{equation}

\subsection{Output of the phase-estimation algorithm}
\label{sec:output}
The output of the phase-estimation algorithm is a series of
measurements of the $w$ number of work qubits. Putting them all together
we get a
probability distribution that estimates the amplitudes $|c_k|^2$ for
each eigenvalue $\phi^{[k]}$.
The $\phi^{[k]}2^w$ 
values we measure from the work qubits, see Eq.~(\ref{eq:FinalState}),  
are  binary numbers from zero to $2^w-1$,
where each one translates to a given eigenvalue of the Hamiltonian depending
on the parameters we have used in our simulation. 
When accurate, a set of simulated
measurements will give a distribution with peaks around the true
eigenvalues. 
The probability
distribution is calculated by applying non-normalized projection
operators to the qubit state,
\[
	\left(| \phi^{[k]}2^t\rangle \langle\phi^{[k]}2^t | \otimes  {\bf 1} \right)
	\left( \sum_{i}c_i |\phi_i 2^t\rangle \otimes |i\rangle \right)
	= c_k|\phi^{[k]} 2^t\rangle \otimes |k\rangle.
\]
The length of this vector squared
gives us the probability,
\begin{equation}
	\left|c_k |\phi^{[k]}\rangle 2^t \otimes |k\rangle \right|^2=|c_k|^2 
	\langle \phi^{[k]} 2^t|\phi^{[k]} 2^t\rangle\langle k |k \rangle= |c_k|^2.
\end{equation} 
Since we do not employ the exact evolution due to different
approximations, we can have non-zero 
probabilities for all values of $\phi$, yielding a distribution
without sharp peaks for the correct eigenvalues and possibly peaks in
the wrong places. If we use different random input states for every
run through the quantum computer and gather
all the measurements in
one probability distribution, all the eigenvectors in the input state
$|\psi\rangle=\sum_k c_k | k\rangle$ should average out to the same amplitude.
This means that eigenvalues with higher multiplicity, i.e., higher
degeneracy, will show up as taller peaks in the probability distribution, 
while non-degenerate eigenvalues might be difficult to find.

To properly estimate the eigenvalues
$E_k$ of the Hamiltonian from this distribution, one must take into account the
periodicity of $e^{2\pi i\phi}$. If $0 < \phi^\prime < 1$ and $\phi =
\phi^\prime +s$, where $s$ is an integer, then
$e^{2\pi i\phi}=e^{2\pi i\phi^\prime}$. This means that to get all
the eigenvalues correctly $\phi$ must be positive and less than one.
Since $\phi = -E_k \Delta t/2\pi$ this means all the eigenvalues $E_k$
must be negative, this merely means subtracting a constant we denote 
$E_{max}$ from the Hamiltonian, $H^\prime = H -E_{max}$, where
$E_{max}$ is greater than the largest eigenvalue of $H$. 
The values we read out from the work qubits are integers from zero to
$2^w-1$. In other words, we have
$\phi^{[k]}2^w\in [0, 2^w-1]$, with $\phi=0$ for $\Delta t=0$. 

The value $\phi=0$ corresponds to the lowest eigenvalue possible to measure,
$E_{min}$, while $\phi=1$ corresponds to $E_{max}$. The interval of
possible values is then $E_{max}-E_{min} = 2\pi/\Delta t$. If we want to
have all possible eigenvalues in the interval the largest value
$\Delta t$ can have is 
\begin{equation}
\mathrm{max}(\Delta t) = \frac{2\pi}{E_{max}-E_{min}}
\end{equation}

\subsubsection{Spectrum analysis}

In the general case one does not know the upper and lower bounds on
the eigenvalues beforehand, and therefore for a given $E_{max}$ and
$\Delta t$ one does not know if the $\phi^{[k]}$ values are the correct
ones, or if an integer has been lost in the exponential function.

When $\phi=\phi^\prime +s$ for one
$\Delta t$, and we slightly change $\Delta t$, $\phi^\prime$ will
change if $s\neq 0$ 
as the period of the exponential function is a function of
$\Delta t$. 
To find out which of $\phi^{[k]}$s are greater than
one, we perform the phase-estimation algorithm with different values for $\Delta
t$ and 
see which eigenvalues shift. 
If we measure the same $\phi$ after adding $\delta t$ to
the time step, and $(\Delta t + \delta t)/\Delta t$ is not a rational
number, we know that $\phi <1$. In practice it does not have to be an
irrational number, but only some unlikely fraction.

There are at least two
methods for finding the eigenvalues. One can start with a
large positive $E_{max}$ and a small $\Delta t$, hoping to find that
the whole spectrum falls within the range $[E_{min}, E_{max}]$, and
from there zoom in until the maximal eigenvalue is slightly less than
$E_{max}$ and the groundstate energy is slightly larger than
$E_{min}$. This way the whole spectrum is covered at once.
From there we can also zoom in on specific areas of the spectrum,
searching  the location of the true eigenvalues by shifting $\Delta t$. 

The number of measurements needed will depend entirely on the
statistics of the probability distribution. The number of eigenvalues
within the given energy range determines the resolution needed. 
That said, the number of
measurements is not a bottleneck in quantum computer calculations. The
quantum computer will prepare the states, apply all the operations
in the circuit and measure. Then it will do it all again. Each
measurement will be independent of the others as the system is
restarted each time. This way the serious problem of decoherence only
apply within each run, and the number of measurements is only
limited by the patience of the scientists operating the quantum
computer.

\section{Results and discussion}
\label{sec:results}
\begin{figure}[h!]
\begin{center}
	\scalebox{0.5}{
		\psfig{file=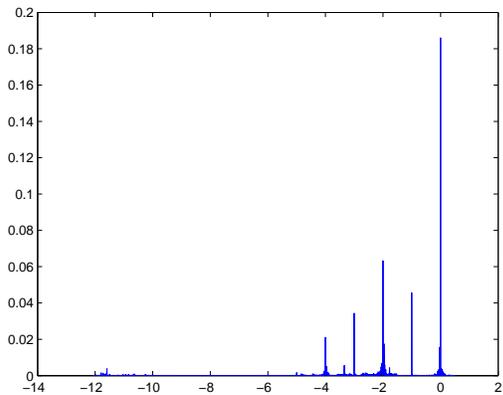}
	}
	\caption{Resulting probability
          distribution from the simulated measurements, giving us the eigenvalues
          of the pairing model with six degenerate energy levels with a total
          possibility of 12  particles and pairing strength $g=1$. 
          The correct eigenvalues are $0$, $-1$, 
          $-2$,  $-3$, $-4$, $-5$, $-6$, $-8$, $ -9$, $-12$.
          All the eigenvalues are not seen as this is the probability
          distribution resulting from one random input state. A
          different random input state in each run could be implemented on
          an actual quantum computer and would eventually yield peaks
          of height corresponding to the degeneracy of each eigenvalue. }
	\label{fig:P24-13-15IM}
\end{center} 
\end{figure}

In this section we present the results for the Hubbard model and the
pairing model of Eqs.~(\ref{eq:hubbard}) and (\ref{eq:pairing}), respectively, and
compare the simulations to exact diagonalization results. 
In Fig.~\ref{fig:P24-13-15IM} we see the resulting probability
distribution from the simulated measurements, giving us the eigenvalues
of the pairing model with six degenerate energy levels and from zero
to  12  particles. 
The pairing strength was set to $g=1$.
The  eigenvalues from the exact solutions of these many-particle
problems are $0$, $-1$, 
$-2$,  $-3$, $-4$, $-5$, $-6$, $-8$, $ -9$, $-12$.
All the eigenvalues are not seen as this is the probability
distribution resulting from one random input state. A
different random input state in each run could be implemented on
an actual quantum computer. These are results for the degenerate
model, where the single-particle energies of the doubly degenerate levels are set to 
zero  for illustrate purposes only, since analytic formula are available for the 
exact eigenvalues. The block diagonal structure of the pairing
Hamiltonian has not been used to our advantage in this straightforward
simulation as the qubit basis includes all particle numbers. 

\begin{figure}[h!]
\begin{center}
	\scalebox{0.5}{
		\psfig{file=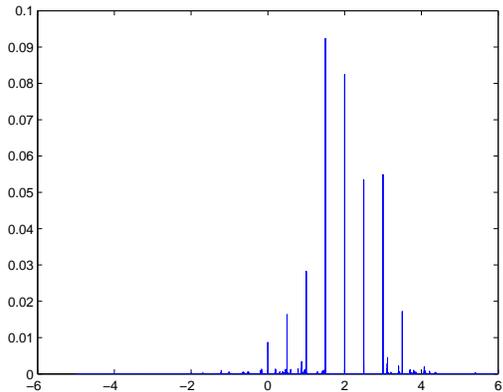}
	}
	\caption{The eigenvalues
          of the non-degenerate pairing model with four energy levels with a total
          possibility of 8  particles, the level spacing $d$ is $0.5$
          and the  pairing strength $g$ is $1$. 
          The correct eigenvalues are obtained from exact
          diagonalization, but in this case there is a multitude of
          eigenvalues and only some eigenvalues are found from this
          first simulation.
        }
	\label{fig:P24-17-e3IMd}
\end{center} 
\end{figure}

 We have also performed tests of the algorithm for
the non-degenerate case, with excellent agreement with our diagonalization codes,
see discussion in Ref.~\cite{rmp75mhj}.
This is seen in Fig.~\ref{fig:P24-17-e3IMd} where we have simulated the pairing model
with four energy levels with a total possibility of eight fermions.
We have chosen $g=1$ and $d=0.5$, so this is a model with low degeneray
and since
the dimension of the system is $2^8= 256$ there is a lot of different
eigenvalues. To find the whole spectrum one would have to employ
some of the techniques discussed in subsection~\ref{sec:output}.

\subsection{Number of work qubits versus number of simulation qubits}

The largest possible amount of different eigenvalues is $2^s$, where
$s$ is the number of simulation qubits. The resolution in the energy
spectrum we get from measuring upon the work qubits is $2^w$, with $w$ the number of
work qubits.
Therefore the resolution per eigenvalue in a non-degenerate system is
$2^{w-s}$. The higher the degeneracy the less work qubits are needed.

\begin{figure}[h!]
\begin{center}
	\scalebox{0.5}{
		\psfig{file=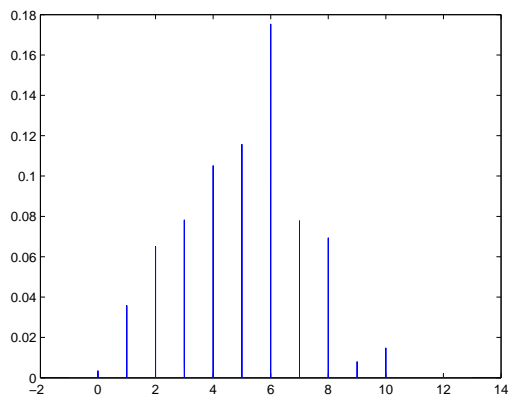}
	}
	\caption{The energy levels of  the Hubbard model of Eq.~(\ref{eq:hubbard}),
          simulated with a total of $24$ qubits, of which eight were
          simulation qubits and $16$ were work qubits.
          In this run  we chose 
          $\epsilon=1$, $t=0$ and $U=1$. The reason we chose $t=0$ was just
          because of the higher degeneracy and therefore fewer
          eigenvalues. The correct eigenvalues are obtained from exact
          diagonalization, with the level of degeneracy following in
          paranthesis: 0(1), 1(8), 2(24), 3(36), 4(40), 5(48), 6(38),
          7(24), 8(24), 9(4), 10(8), 12(1).  
        }
	\label{fig:24-17-1T0}
\end{center} 
\end{figure}

In Fig.~\ref{fig:24-17-1T0} we see the results for the Hubbard model
Eq.~(\ref{eq:hubbard})
with $\epsilon=1$, $t=0$ and $U=1$. The reason we chose $t=0$ was just
because of the higher degeneracy and therefore fewer eigenvalues.
 The number of work qubits is $16$
and the number of simulation qubits is eight for a total of $24$
qubits. The difference between work qubits and simulation qubits is
eight which means there are $2^8$ possible energy values for each
eigenvalue. Combining that with the high degeneracy we get a very sharp
resolution. The correct eigenvalues with degeneracies are obtained from exact
diagonalization of  the Hamiltonian, the degeneracy follows the
eigenvalue in paranthesis: 0(1), 1(8), 2(24), 3(36), 4(40), 5(48), 6(38),
7(24), 8(24), 9(4), 10(8), 12(1). We can clearly see that even though
we have a random input state, with a random superposition of the
eigenvectors, there is a correspondence between the height of the
peaks in the plot and the degeneracy of the eigenvalues they represent.

\subsection{Number of time intervals}

The number of time intervals, $I$, is the number of times we must 
implement the time evolution operator in order to reduce the error in the
Trotter approximation  \cite{trotter1959,suzukitrotter,suzuki1985}, 
see Eq.~(\ref{eq:Trotter1}).
In our program we have only implemented the simplest Trotter
approximation and in our case we find that we do not need a large $I$
before the error is small enough. In Fig.~\ref{fig:24-17-1T0} 
$I$ is only one, but here we have a large number of work qubits.
For other or larger systems it
might pay off to use a higher order Trotter approximation. The total
number of operations that have to be done is a multiple of $I$, but 
this number
also increases for higher order Trotter approximations, so for each
case there is an optimal choice of approximation.

\begin{figure}[h!]
\begin{center}
	\scalebox{0.5}{
		\psfig{file=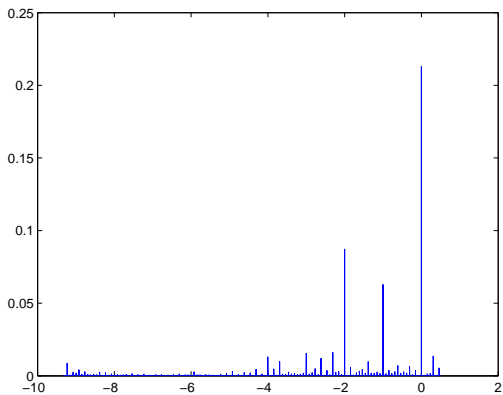}
	}
	\caption{Pairing model simulated with $24$ qubits, where $14$
          were simulation qubits, i.e. there are $14$ available
          quantum levels,  and
          $10$ were work qubits.  The correct eigenvalues are $0$, $-1$,
          $-2$,  $-3$, $-4$, $-5$, $-6$, $-8$, $ -9$. In this run we
          did not divide up the time interval to reduce the error in
          the Trotter approximation, i.e., $I=1$.}
	\label{fig:P24-15-1IM}
\end{center} 
\end{figure}

\begin{figure}[h!]
\begin{center}
	\scalebox{0.5}{
		\psfig{file=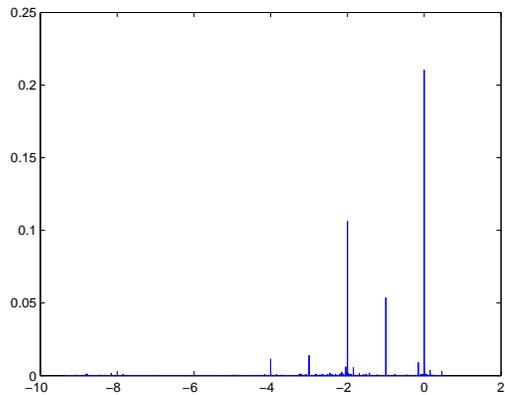}
	}
	\caption{Pairing model simulated with $24$ qubits, where $14$
          were simulation qubits, i.e. there are $14$ available
          quantum levels,  and
          $10$ were work qubits.
          The correct eigenvalues are $0$, $-1$,
          $-2$,  $-3$, $-4$, $-5$, $-6$, $-8$, $ -9$. In this run we
          divided the time interval into $10$ equally space parts
          in order to reduce the error in
          the Trotter approximation, i.e., $I=10$.}
	\label{fig:P24-15-e1IM}
\end{center} 
\end{figure}

In Figs.~\ref{fig:P24-15-1IM} and \ref{fig:P24-15-e1IM} we 
see the errors deriving from the Trotter approximation, and how they
are reduced by increasing the number of time intervals. The results in this figure are
for the degenerate pairing model  with 24 qubits in total, and ten
simulation qubits with $d=0$ and $g=1$. In Fig.~\ref{fig:P24-15-1IM}
we had $I=1$ while in Fig.~\ref{fig:P24-15-e1IM} $I$ was set to ten. Both
simulations used the same starting state. The errors are seen as the
small spikes around the large ones which represent some of the
eigenvalues of the system. The exact eigenvalues are  $0$, $-1$,
$-2$,  $-3$, $-4$, $-5$, $-6$, $-8$, $ -9$. 

\subsection{Number of operations}

Counting the number of single-qubit and $\sigma_z\sigma_z$ operations
for different sizes of systems simulated gives us an indication of the
decoherence time needed for different physical realizations of a
quantum simulator or computer. The decoherence time is an average time
in which the state of the qubits will be destroyed by noise, also called 
decoherence, while the operation time is the average time an operation takes
to perform on the given system. Their fraction is the number of
operations possible to perform before decoherence destroys the
computation. In table~\ref{fig:noOfGates} we have listed the number of
gates used for the pairing model, $H_P$, and the Hubbard model, $H_H$,
for different number of simulation qubits. 
\begin{table}[h!]
\begin{center}
\begin{tabular}{l|cccccc}\hline\hline
	 & $s=2$&$s=4$&$s=6$&$s=8$&$s=10$ & $s=12$ \\ \hline
	 $H_P$ & 9 & 119 & 333 & 651 & 1073 & 1598 \\ \hline
         $H_H$ & 9 &  51 &  93 & 135 & 177  & 219  \\ \hline
\end{tabular}
\end{center}
\caption{Number of two-qubit gates used in simulating the time
  evolution operator of the pairing model, $H_P$, and the Hubbard
  model, $H_H$, for different number of simulation qubits $s$.}
\label{fig:noOfGates}
\end{table}

\section{Conclusion}
\label{sec:conclusion}
In this article we have shown explicitly how the Jordan-Wigner
transformation is used to simulate any many-body fermionic Hamiltonian
by two-qubit operations. We have shown how the simulation of such
Hamiltonian terms of products of creation and annihilation operators
are represented by a number of operations linear in the number of
qubits. 
To perform efficient quantum simulations on quantum computers one
needs transformations that take the Hamiltonian in question to a set
of operations on the qubits simulating the physical system. An example of 
such a transformation employed in ths work, is 
the Jordan-Wigner transformation.  With the appropriate transformation and
relevant gates or quantum circuits, one can taylor  an actual quantum computer to
simulate and solve the eigenvalue and eigenvector problems for different quantum systems. 
Specialized quantum simulators might be more efficient in solving some
problems than others because of similarities in algebras between physical system
of qubits and the physical system simulated. 

We have limited the applications to two simple and well-studied models that provide,
via exact eigenvalues, a good testing ground for our quantum computing based 
algorithm.  For both the pairing model and the Hubbard model we obtain an excellent agreement.
We plan to extend the area of application to quantum mechanical studies  of systems in nuclear physics,
such as a comparison of shell-model studies of oxygen or calcium isotopes where the nucleons are active 
in a given number of single-particle orbits \cite{mhj95,caurier2005}. These single-particle orbits have normally a higher 
degeneracy than $2$, a degeneracy  studied here.  However, the algorithm we have developed allows
for the inclusion of any  degeneracy, meaning in turn that with a given interaction $V_{ijkl}$   
and single-particle energies, we can compare the nuclear shell-model (configuration interaction) calculations
with our algorithm.

\section*{Acknowledgment}
This work has received support from the Research Council of Norway through the center of
excellence program.

\appendix*

\section{Useful relations}

We list here some useful relations involving different $\sigma$ matrices,
\begin{equation}
\sigma_x \sigma_z = -i\sigma_y, \quad
\sigma_z \sigma_x = i\sigma_y, \quad [\sigma_x, \sigma_z]=-2i\sigma_y,
\end{equation}
\begin{equation}
\sigma_x \sigma_y = i\sigma_z, \quad
\sigma_y \sigma_x = -i\sigma_z, \quad [\sigma_x, \sigma_y]=2i\sigma_z,
\end{equation}
and
\begin{equation}
\sigma_y \sigma_z = i\sigma_x, \quad
\sigma_z \sigma_y = -i\sigma_x, \quad [\sigma_y, \sigma_z]=2i\sigma_x.
\end{equation}

For any two non-equal $\sigma$-matrices $a$ and $b$ we have 
\begin{equation}
aba = -b.
\end{equation}

The Hermitian $\sigma$-matrices $\sigma_x$, $\sigma_y$ and $\sigma_z$
result in the identity matrix when squared
\begin{equation}
\sigma_x^2 = {\bf 1},\quad 
\sigma_y^2 = {\bf 1},\quad 
\sigma_z^2 = {\bf 1},\quad 
\end{equation}
which can be used to obtain  simplified expressions for exponential functions involving $\sigma$-matrices
\begin{equation}
e^{\pm i\alpha \sigma}=\cos(\alpha) {\bf 1} \pm i \sin(\alpha) \sigma. 
\end{equation}

The equations we list below are necessary for the relation between  a general unitary
transformation on a set of qubits with  a product of two-qubit unitary
transformations. We have the general equation for $a,b \in \{\sigma_x,\sigma_y, \sigma_z\}$, where $a\neq b$.
\begin{eqnarray}
  e^{-i\pi/4a} b e^{i\pi/4a} &&= \frac{1}{2} ({\bf 1} -ia) b ( {\bf 1} + ia)
  \notag\\ 
  &&=  \frac{1}{2} (b + aba + i[b,a]) \notag\\
  &&= \frac{i}{2}[b,a].
\end{eqnarray}
The more specialized equations read
\begin{eqnarray}
  \label{eq:rotations1}
  &&e^{-i\pi/4 \sigma_x} \sigma_z e^{i\pi/4 \sigma_x} = -\sigma_y, \\
  \label{eq:rotations2}
  &&e^{-i\pi/4 \sigma_y} \sigma_z e^{i\pi/4 \sigma_y} = \sigma_x, \\
  \label{eq:rotations3}
  &&e^{-i\pi/4 \sigma_z} \sigma_x e^{i\pi/4 \sigma_z} = \sigma_y, \\
  \label{eq:rotations4}
  &&e^{-i\pi/4 \sigma_z} \sigma_y e^{i\pi/4 \sigma_z} = -\sigma_x. 
\end{eqnarray}

We need also different products of the operator$\sigma_z$  with the  raising and lowering operators
\begin{eqnarray}
\label{eq:pmzs}
  &&\sigma_+ \sigma_z = -\sigma_+ \\
  &&\sigma_z \sigma_+ = \sigma_+, \\
  &&\sigma_- \sigma_z = \sigma_-, \\
  &&\sigma_z \sigma_- = -\sigma_-. \\
\end{eqnarray}

\bibliographystyle{unsrt}

\end{document}